\documentclass[journal]{IEEEtran}
\usepackage{graphics} 
\usepackage{epsfig} 
\usepackage{mathptmx} 
\usepackage{times} 
\usepackage{amsmath} 
\usepackage{amssymb}  
\usepackage{enumitem}
\usepackage{array}
\usepackage{ulem}
\usepackage{cite}
\usepackage{multirow}
\usepackage{booktabs}

\usepackage{algorithm}
\usepackage{algorithmicx}
\usepackage[noend]{algpseudocode}
\usepackage{hyperref}

\usepackage{xcolor}
\usepackage[dvipsnames]{xcolor}

\newcommand{\Rf}[1]{\textcolor{black}{#1}}
\newenvironment{rf}{\color{black}}{}

\begin{document}

\title{Explainable Control Framework (XCF) based on Fuzzy Model-Agnostic Explanation and LLM Agent-Supported Interface}

\author{Faliang Yin, Hak-Keung Lam,~\IEEEmembership{Fellow,~IEEE}, David Watson
\vspace{-1em}
\thanks{This research is partially supported by King’s-China Scholarship Council PhD Scholarship programme. \textit{(Corresponding author: Hak-Keung Lam.)}}
\thanks{Faliang Yin and  Hak-Keung Lam are with the Department of Engineering,
Faculty of Natural, Mathematical \& Engineering Sciences, King's College London, WC2R 2LS London, United Kingdom (e-mail: faliang.yin@kcl.ac.uk; hak-keung.lam@kcl.ac.uk).}
\thanks{David Watson is with the Department of Informatics, Faculty of Natural, Mathematical \& Engineering Sciences, King's College London, WC2R 2LS London, United Kingdom (e-mail: david.watson@kcl.ac.uk).}
}

\markboth{Preprint submitted to IEEE for possible publication.}%
{Shell \MakeLowercase{\textit{et al.}}: Bare Demo of IEEEtran.cls for IEEE Journals}

\maketitle

\begin{center} \begin{minipage}{0.95\linewidth} \small \textbf{Preprint Notice} This work has been submitted to the IEEE for possible publication. Copyright may be transferred without notice, after which this version may no longer be accessible. \end{minipage} \end{center}

\begin{abstract}
Increasing demand for precise and reliable control in complex scenarios has led to the development of increasingly sophisticated controllers, including data-driven approaches employing closed box models and mathematically rigorous yet complex designs. This complexity highlights the needs for explainable control that can provide human-understandable insights into controller behavior. In this paper, an explainable control framework (XCF) along with supporting algorithms and user interface are proposed to explain how controllers determine their control actions and their underlying working mechanism. The novel contributions of this work are threefold: First, the XCF is designed to provide model-agnostic explanations for controllers in closed-loop systems and can optionally refine local explanations by system response dynamics. Second, a novel explanation method, hierarchical fuzzy model-agnostic explanation for control systems (HFMAE-C), is proposed based on the designed framework. The HFMAE-C employs a fuzzy logic system to approximate the controller's behavior and system dynamics, providing sample, local, domain and universe level explanations via IF-THEN rules revealing the controller's decision logic and salience values quantifying the contribution of system states to control actions. Third, a large language model agent-supported user interface is developed to automatically analyze user requirements, select appropriate algorithms, interpret the generated explanations to a natural language report, and provide interactive consultation. \Rf{Case studies on inverted pendulum system and Turtlebot obstacle avoidance demonstrate the effectiveness of the proposed method through simulated user experiments and quantitative comparisons with mainstream explainable control approaches.} 
\end{abstract}


\IEEEpeerreviewmaketitle

\section{Introduction}
\IEEEPARstart{T}{o} achieve high performance, adaptability, and automation, modern control systems increasingly rely on sophisticated models and advanced control strategies. From traditional model-based control to data-driven approaches, the complexity of controllers grows significantly, making their internal decision-making process more difficult to interpret. On the one hand, many high-performance controllers are designed based on rigorous control theories, such as nonlinear control including fuzzy model-based control \cite{case}, sliding mode control, and backstepping control, yet their mathematical complexity often limits accessibility to engineers and operators. 

On the other hand, machine learning control methods \cite{overview} learn the mapping between system states and control actions directly from data, which often results in closed-box controllers and consequently raises new challenges for explainability. Despite this limitation, such data-driven approaches have been increasingly adopted in control systems due to their strong adaptability and flexibility. For example, convolutional neural networks are widely used in vision-based control tasks, such as search-and-rescue quadcopters \cite{CNN1} that predict trail directions to adjust motion, and underwater snake robots \cite{CNN2} that perform targeted docking in unstructured environments. Recurrent neural networks have been integrated into bipedal robot controllers \cite{RNN} to approximate static and dynamic friction, while deep neural networks have been employed as model predictive controllers for quadcopters \cite{DNN}. In addition, reinforcement learning (RL) \cite{rein} has been widely adopted in control applications, enabling agents to learn optimal policies through trial-and-error interactions with the environment. A recent review \cite{overview} provides a comprehensive summary of machine learning-based control methods.

As control systems become increasingly sophisticated, explainability becomes essential for understanding, debugging, and trusting both data-driven and theoretically grounded yet complex control methods. In machine learning control, where controllers are learned from data rather than explicitly defined models, the internal logic remains opaque, raising concerns in high-stakes applications such as industrial automation \cite{cyber1,cyber2}. Meanwhile, mathematically rigorous control strategies, despite being built on well-established theoretical principles, often involve complex computations and optimizations that make them challenging for engineers and operators to intuitively understand and analyze \cite{flight, av, ilc}. 

This challenge of explainability mirrors a similar problem faced in the artificial intelligence (AI) field, where powerful but opaque closed box models raise concerns about trust and reliability. To explain model decisions, the AI community has developed post-hoc explainable AI (XAI) \cite{survey_scale}, a collection of methods and tools to generate explanations after model training. Among these, model-agnostic methods are widely used, as they generate explanations solely based on model inputs and outputs without requiring access to the model's internal structure. Representative methods include local interpretable model-agnostic explanations (LIME) \cite{LIME}, which explains individual predictions by learning a local linear approximation of the model; Shapley Additive explanations (SHAP) \cite{SHAP}, which attributes feature contributions based on principles from cooperative game theory; and fuzzy model-agnostic explanation (FMAE) \cite{FMAE, FMAE2}, which provides hierarchical explanations for global to local model behavior by feature salience values and IF-THEN fuzzy rules. 
Despite clear differences between these methods, they all aim to return (approximately) minimal conditions sufficient to alter predictions in some prespecified way \cite{watson_lens, watson_conceptual}. 
Resulting explanations should be accurate, simple, and relevant for the inquiring agent \cite{watson_exp_game}. 
These techniques have been widely applied to domains such as computer vision and natural language processing, and hold potential for improving explainability in control systems.

Inspired by recent advances in XAI, explainable control has emerged as an active research topic for increasingly complex control systems. Various studies have explored integrating XAI techniques into different control scenarios. For instance, a hybrid LIME–SHAP framework has been applied to autonomous vehicles to provide real-time explanations of control decisions \cite{av}. SHAP has also been introduced into iterative learning control to improve the transparency of control parameter effects \cite{ilc}. In control system fault diagnosis, causal language models have been used to support explainable cause–effect reasoning \cite{llm}. An XAI framework has been developed for real-time control software, where SHAP and LIME are employed to identify influential features in defect prediction \cite{software}. Explainability has been increasingly integrated into RL-based control, with applications reported in marine docking \cite{docking}, vehicle powertrain systems \cite{powertrain}, power system emergency response \cite{emergency}, and smart building energy optimization \cite{energyopt}. Moreover, recent studies also investigated the use of large language models (LLMs) in explainable control for human-understandable contexts \cite{llm2, llm3}.

Despite the encouraging progress made in explainable control, existing methods still face two key challenges that limit their broader applicability: universality and accessibility. Universality refers to the ability of a method to generalize across different types of control systems and application scenarios, rather than being tailored to a specific domain or controller architecture. Many existing approaches are tightly coupled with particular system structures, making it difficult to transfer or reuse them in other control contexts. Accessibility concerns the ease with which a user (especially one without deep expertise in control theory or XAI) can apply the method and interpret the generated explanations \cite{llm2, llm3}. Current solutions often require detailed knowledge of both the control mechanisms and XAI methods, posing a barrier for domain experts and practitioners seeking intuitive insights. 

Existing model-agnostic XAI methods mainly explain static input-output relationships. In closed-loop control systems, however, controller behavior is associated with dynamic system responses through feedback interactions. Therefore, approximation quality based only on controller input-output mapping may not fully reflect behavioral consistency between the explainer and the original controller. 
To address these limitations, this paper proposes an explainable control framework (XCF) together with supporting explanation algorithms and a user interface to provide explainability with universality and accessibility for closed-loop control systems. Our recent conference publication \cite{xcf_c} introduced preliminary concepts of the XCF. In this paper, the framework is substantially developed through new methodological, algorithmic, interface, and experimental contributions. In particular, beyond conventional static input-output approximation, closed-loop response dynamics are optionally incorporated into the surrogate modeling process to improve the behavioral fidelity between the explainer and the original controller. The main novelties of this journal paper are summarized as follows:

\begin{rf}
\begin{enumerate}[leftmargin=*]
\item An XCF is developed as a general model-agnostic framework to explain the behavior of controllers in closed-loop systems. Closed-loop response information is incorporated into the local surrogate modeling and refinement process to improve the behavioral fidelity of the explainer with respect to the original controller, where behavioral fidelity refers to the degree to which the explainer not only approximates the controller input-output mapping, but induces closed-loop responses consistent with those generated by the controller.

\item The hierarchical FMAE for control systems (HFMAE-C) is proposed to implement the XCF, providing sample-, local-, domain-, and universe-level explanations through IF-THEN rules and state salience values. A novel three-step algorithm is proposed for local explainer learning and refinement, where both controller input-output relationships and closed-loop response dynamics are considered in the surrogate modeling. Also, a rule aggregation algorithm is developed to construct structured domain explainers through input-dependent aggregation of local explainers.

\item A user interface supported by an LLM agent system is developed to automatically identify user explanation intent, select suitable explanation algorithms, and translate structured fuzzy explanations, including IF-THEN rules and state salience values, into accessible natural-language reports and interactive consultation.

\item Inverted pendulum and Turtlebot obstacle avoidance case studies are presented to evaluate the proposed XCF on both intuitive and complex control scenarios. The inverted pendulum case demonstrates hierarchical explanations and LLM-assisted interpretation, and the Turtlebot case provides quantitative comparison with widely adopted model-agnostic explanation methods, including SHAP and LIME.
\end{enumerate}
\end{rf}

\noindent \textbf{Remark} The term \textit{controller} denotes a state-to-action mapping in control systems. Unless otherwise specified, all references to the controller correspond to the original closed box controller $u(\mathbf{x})$. The explainer is always explicitly denoted by $\hat{u}(\mathbf{x})$.

\section{Explainable Control Framework (XCF)}

\begin{figure}[htbp]
\centering
\includegraphics[width=0.5\textwidth]{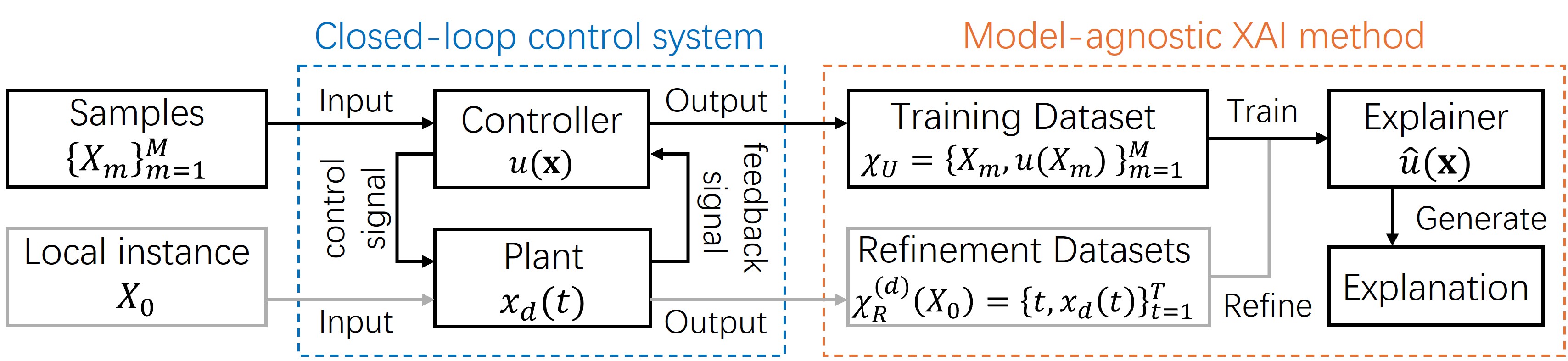}
\caption{Diagram of explainable control framework (XCF). The black blocks and gray blocks represent necessary and optional parts, respectively.}
\label{fig:XCF}
\end{figure}

To address the universality challenge in explainable control, we propose the XCF employing model-agnostic XAI methods to explain the controller in a closed-loop control system. XCF is designed to be universal and applicable to various types of controllers, including closed box data-driven models and complex theoretical control strategies. Unlike existing explainability approaches tailored to specific problems, XCF provides a systematic method to extract explainable insights without requiring access to the controller's internal structure. As shown in Fig. \ref{fig:XCF}, the framework consists of two key components: a closed-loop control system, where a controller $u(\mathbf{x})$, with the system state vector $\mathbf{x}=[x_1, x_2, \ldots, x_D]$, interacts with a plant through feedback signals, and a model-agnostic XAI method, which approximates and explains the controller’s behavior. 
Optionally, XCF can incorporate information from the closed-loop system response to locally refine the explainer. 
In Fig. \ref{fig:XCF}, the black blocks and gray blocks represent the necessary and optional parts, respectively.

XCF begins by sampling system states within a region of interest specified by the user. For example, if the user expects to analyze the controller’s global behavior, sampling is conducted over the entire value range of system states; if the user is interested in the local decision around a specific system state $X_0$, sampling is restricted to a neighborhood around that point. Each sampled state $X_m$ is passed through the controller $u(\mathbf{x})$, and the resulting control signals $u(X_m)$ form a dataset $\{(X_m, u(X_m))\}_{m=1}^M$. This dataset is then used to train a model-agnostic explainer $\hat{u}(\mathbf{x})$, which aims to approximate the input-output behavior of the original controller. The learned explainer is subsequently used to generate human-understandable explanations of the controller's behavior.

Optionally, if the user focuses on the controller's local decision near a particular state value $X_0$, XCF supports a refinement step that utilizes the closed-loop control system's feedback dynamics to enhance the local approximation accuracy of the explainer. Specifically, by solving the system’s differential equations under both the original controller $u(\mathbf{x})$ and the explainer $\hat{u}(\mathbf{x})$, we obtain the corresponding responses of the $d$th state $x_d(t)$ and $\hat{x}_d(t)$, respectively. The response $x_d(t)$ from the controller, together with the time variable $t$, can be used to construct a refinement dataset $\{(t, x_d(t))\}_{t=1}^T$, $t=1, 2, ..., T$. The difference between the responses $x_d(t)$ and $\hat{x}_d(t)$ are defined as the response error quantifying the behavioral divergence introduced by the approximation. By minimizing the response error (e.g., incorporate it into loss function) in this localized region, the explainer can be refined to more accurately reflect the controller’s behavior around $X_0$.

\section{Hierarchical Fuzzy Model-Agnostic Explanation for Control systems (HFMAE-C)}
HFMAE-C is proposed as an explanation algorithm universally capable of interpreting the behavior of a given controller in a closed-loop system. This section introduces the hierarchical explanation structure, which organizes explanations into sample, local, domain, and universe levels; the corresponding model-agnostic explanation algorithms, which train explainers for each level; and the explanation interface of the trained explainer, which provides salience values for system states and IF-THEN rules revealing the controller’s decision logic.

\subsection{Hierarchical Explanation Structure}
HFMAE-C organizes the explanations of controller behavior into a structured hierarchy consisting of four levels: sample, local, domain, and universe. Each level corresponds to a different scale of analysis over the controller behavior. The universe includes all domains, a domain includes multiple localities, and a locality includes numerous samples. The sample explanation captures the controller decision at a specific system state. The local explanation summarizes the controller behavior within a small neighborhood around a given system state instance. The domain explanation generalizes over a larger region of the state space where similar behaviors are observed, typically formed by aggregating several local explanations. The universe explanation provides a global summary of the controller decision logic across the entire state space. 

As shown in the existing work on FMAE \cite{FMAE}, such hierarchical structures allow both bottom-up aggregation and top-down simplification of explanations, and users can begin analysis at any level based on their needs. In the following sections, we describe how the explainers for different levels are constructed in a practical implementation: a univers explainer is trained directly from global data; local explainers are trained and refined for given state instances; and domain explainers are obtained by aggregating multiple local explainers.

\subsection{Model-agnostic Explanation Algorithms for Control System}
To generate hierarchical explanations in a model-agnostic manner, HFMAE-C employs fuzzy logic system (FLS) as surrogate models to approximate the input-output behavior of the target controller. This section first introduces the structure of the FLS-based explainer, followed by the construction of explainers at the universe, local, and domain levels.

\subsubsection{FLS-Based Explainer}
FLS is adopted as the explainer to approximate controller behavior due to its proven advantages \cite{FMAE} in transferability for sharing knowledge between agents, scalability for providing hierarchical explanations, and comprehensibility for humans to understand. In particular, transferability plays a unique role in this work by enabling natural language rule-based explanations to be effectively interpreted by LLMs in next section.

The FLS $\hat{u}(\mathbf{x})$ employed as the explainer is formed by a fully combined fuzzy rule base $R=\{r_1,r_2,\ldots,r_N\}$. A typical linguistic form of a TSK \cite{TSK} fuzzy rule $r_n$ consists of an antecedent (IF part) and a consequent (THEN part):
\begin{align}
r_n: ~&IF\ x_1\ is\ A^1_n\ and\ ...\ and\ x_D\ is\ A^D_n,\nonumber \\
&THEN\ f_{n}(\mathbf{x}) =  a^0_{n}+a_{n}^1x_1+a_{n}^2x_2+...+a_{n}^Dx_D. 
\label{eq:fuzzymodel_rule_n}
\end{align}

In the antecedent, $x_d$ denotes the $d$th system state and $A^d_n$ denotes the corresponding fuzzy set in the $n$th rule, where $d=1,2,\ldots,D$ and $n=1,2,\ldots,N$. For each premise such as ``$x_d\ is\ A^d_n$'', the degree of truth is measured by a membership function. In the consequent, $f_n(\mathbf{x})$ is a trainable linear model, where $a^d_n$ denotes the scalar coefficient associated with the $d$th system state and $a^0_n$ denotes a scalar bias. The normalized firing strength of the $n$th rule is denoted by $\omega_n(\mathbf{x})$, representing its activation level for the input state $\mathbf{x}$. 
The output of the FLS is computed as the weighted average of all fuzzy rules:
\begin{equation}
\hat{u}(\mathbf{x})=\sum^N_{n=1}\omega_n(\mathbf{x}) ~f_{n}(\mathbf{x}).
\label{out}
\end{equation}

HFMAE-C employs the above FLS $\hat{u}(\mathbf{x})$ as the fundamental explainer used throughout all explanation levels. The complete formulation of the FLS is provided in the \textit{Supplementary Materials}. For different levels, explainers can be constructed by adjusting the sampling strategy and learning objective. In the following three parts, we introduce the algorithms for training hierarchical explainers to capture controller behavior at varying levels of granularity.

\subsubsection{Universe explainer}
The universe explainer aims to approximate the controller $u(\mathbf{x})$ over the full admissible range of the system state. Let the universe training dataset be $\chi_U=\{(X_m, u(X_m))\}_{m=1}^{M}$, where the samples $X_m$ are drawn across the entire state space. The universe explainer $\hat{u}_{\mathrm{U}}(\mathbf{x})$ is trained by minimizing the global prediction error
\begin{equation}
L_{U}
=\sum_{m=1}^{M}\bigl(u(X_m)-\hat{u}_{\mathrm{U}}(X_m)\bigr)^2 .
\label{eq:U_total}
\end{equation}
This optimization can be performed by gradient-based methods, yielding a global approximation of the controller behavior. The trained universe explainer serves as a reference model when constructing and refining more specialized explainers at the domain and local levels.

\subsubsection{Local explainer}
A local explainer aims to approximate the behavior of the controller associated with a specific initial state $X_0$ to provide local explanations of the $X_0$ neighborhood and sample explanations within this range. Because the system trajectory starting from $X_0$ may pass through regions extending beyond its neighborhood, the learning procedure combines local-scale learning, global-assisted, and response-assisted refinement. This structure enables the explainer to capture both the local behavior and the dynamic behavior of the controller along the entire convergence path. 
Let the local dataset sampled around instance $X_0$ be 
$\chi_L(X_0)=\{(X_k, u(X_k))\}_{k=1}^{K}$, 
and for each state dimension $d$, let the response refinement dataset be 
$\chi_R^{(d)}(X_0)=\{(t, x_d(t))\}_{t=1}^{T}$, 
where $x_d(t)$ denotes the $d$th component of the trajectory generated by the original controller from $X_0$. The training consists of three steps.

\textbf{Step 1:} Local-scale learning. 
The first step focuses on approximating the controller in the neighborhood of the initial state $X_0$. The loss function is
\begin{equation}
L_{\mathrm{S1}}
=
L_{\mathrm{pre}}\bigl(\chi_L(X_0)\bigr)
=
\sum_{X_k\in\chi_L(X_0)}
\bigl(u(X_k)-\hat{u}_{\mathrm{loc}}(X_k)\bigr)^2 .
\label{eq:local_s1}
\end{equation}
This stage is trained with a dedicated number of iterations (e.g., 2000) to obtain a locally accurate initialization.

\textbf{Step 2:} Global-assisted refinement. 
Although the explainer is local by design, the trajectory beginning at $X_0$ may leave its neighborhood during the convergence process. To ensure that the explainer remains effective outside the local region, refinement is performed by incorporating the universe dataset $\chi_U$. The objective is a convex combination of the local and universe prediction losses:
\begin{equation}
L_{\mathrm{S2}}
=
(1-\lambda_{u})\,L_{\mathrm{pre}}\bigl(\chi_L(X_0)\bigr)
+\lambda_{u}\,L_{\mathrm{pre}}\bigl(\chi_U\bigr),
\label{eq:local_s2}
\end{equation}
where $\lambda_u\in[0,1]$ determines the contribution of the universe data.
Only a small number of iterations (e.g. 500) are used, since the goal is to improve general consistency without overriding the local behavior captured in Step 1.

\textbf{Step 3:} Response-assisted refinement. 
To improve the approximation of the system’s actual response from the initial state $X_0$ the explainer is further refined by comparing closed-loop trajectories. Let 
$\mathbf{x}(t;X_0)$ and $\hat{\mathbf{x}}(t;X_0)$ 
denote the trajectories produced by the original controller and the explainer, respectively. For each dimension $d$, define the response error:
\begin{equation}
L_{\mathrm{res}}\bigl(\chi_R^{(d)}(X_0)\bigr)
=
\sum_{t=1}^{T}
\bigl(x_d(t;X_0)-\hat{x}_d(t;X_0)\bigr)^2 .
\label{eq:local_response}
\end{equation}

The refinement objective is
\begin{equation}
L_{\mathrm{S3}}
=
\Bigl(1-\sum_d\lambda_d\Bigr)
L_{\mathrm{pre}}\bigl(\chi_L(X_0)\bigr)
+
\sum_{d}\lambda_d\,
L_{\mathrm{res}}\bigl(\chi_R^{(d)}(X_0)\bigr),
\label{eq:local_s3}
\end{equation}
where $\lambda_d\ge0$ is the contribution of each state dimension. 
Since computing the response errors involves numerical integration of the system dynamics, this step uses only a small number of iterations (e.g. 50) but significantly improves the alignment between the trajectories induced by explainer and controller.

\subsubsection{Domain explainer}

A domain explainer is constructed to approximate the controller’s behavior over a region of the state space in which multiple local explainers exhibit similar patterns. Suppose a set of local explainers $\{\hat{u}_1(x), \hat{u}_2(x), \ldots, \hat{u}_S(x)\}$ has been obtained for several initial states within the region of interest. The aim of domain explainer is to aggregate these local explainers, by weight aggregation or rule aggregation, into a single model that represents their collective behavior.

\textbf{Weight aggregation}. 
Existing work \cite{FMAE} adopts weight strategy, where the domain explainer is expressed as a linear combination of the local explainers. Given a domain dataset $\chi_D$, the aggregated output takes the form
\begin{equation}
    \hat{u}_A(x) = w^0 + w^1\hat{u}_1(x) + w^2\hat{u}_2(x) + \dots + w^S\hat{u}_S(x),
\end{equation}
with the coefficients $w^0, w^1, \ldots, w^S$ computed through least-squares regression on $\chi_D$. This approach provides a straightforward means of combining local explainers and yields a single set of weights that applies across the entire domain.

\textbf{Rule aggregation}. Although weight aggregation is computationally efficient, its expressiveness is limited because the same weight vector applies to all inputs. Consequently, it is unable to reflect variations in the domain level behavior that arise from state-dependent differences in the contributions of the local explainers. To address this limitation, we propose a rule aggregation algorithm that integrates the local explainers within the structure of a fully combined fuzzy rule base, thereby enabling input-dependent aggregation. In the proposed framework, a set of fuzzy rules is constructed using the same antecedent structure as a typical FLS. Each rule partitions the domain through fuzzy sets, and its consequent is defined as a weighted combination of the outputs of all local explainers:
\begin{align}
\text{IF } x_1 \text{ is } A_n^1 \text{ and }\; \ldots \text{ and }\; x_D \text{ is } A_n^D, \text{ THEN }\nonumber \\
     f_{n}(\mathbf{x})= w_n^0 + w_n^1\hat{u}_1(x) + w_n^2\hat{u}_2(x) + \dots + w_n^S\hat{u}_S(x).
\end{align}
The coefficients $w_n^0, w_n^1, \ldots, w_n^S$ are obtained by solving a regularized least-squares problem on the samples falling within the domain. As a result, the contribution of each local explainer varies across different regions of the domain according to both the fuzzy memberships in the antecedent part and the rule-specific coefficients in the consequent part.

By embedding the local explainers into the fuzzy rule consequents, rule aggregation enables a finer-grained representation of the domain level behavior. The aggregated output is obtained by weighting all rule consequents according to their firing strengths, leading to input-dependent combinations that cannot be expressed through a single weight vector. Compared with weight aggregation, the proposed method provides greater flexibility and captures structured variations within the domain, while maintaining compatibility with the fuzzy rule form required by the hierarchical explanation structure.

\subsection{Explanation Interface}

The explanation interface provides structured information that describes how the control output is produced, which can later be translated by the LLM-based user interface introduced in the next section. By using intrinsically interpretable quantities in FLS, such as membership degrees and rule firing strengths, the interface improves the transparency of control decisions and provides richer evidence for detailed explanations. The explanation interface contains two components: state salience, which quantifies the influence of each system state on the control output, and semantic inference, which summarizes the rule-level decision tendencies of the rule base.

\textbf{State salience} measures the contribution of each system state based on the consequent parameters and the activation patterns of the fuzzy rules. For a rule base with $N$ rules, let $a_n^d$ denote the consequent coefficient associated with the $d$th state in rule $n$, and let $\omega_n(X_m)$ denote the normalized firing strength of rule $n$ for sample $X_m$ ($m=1,2,...,M$). The salience value for the $d$th system state is computed as
\begin{equation}
    \bar{\alpha}^d = \frac{1}{M} \sum_{m=1}^M \sum_{n=1}^N \omega_n(X_m)\,a_n^d.
\end{equation}
By assigning a salience value to each state, it offers a compact description of how individual states influence the control output. A positive salience value indicates that the corresponding state contributes positively to the control action, while a negative value indicates a suppressing effect. 

\textbf{Semantic inference} summarizes the behavior of the fuzzy rule base by evaluating each rule’s representative decision. Although the antecedent part already provides a linguistic description of the input conditions, the consequent part is a linear system and may not be sufficiently direct for explanatory purposes. To simplify the representation, the decision tendency of rule $r_n$ is summarized by a decision score defined as the weighted average of the rule’s predicted outputs:
\begin{equation}
    \bar{f}_{n} = \frac{\sum_{m=1}^{M} \omega_n(X_m)\,\hat{u}(X_m)}{\sum_{m=1}^{M} \omega_n(X_m)}.
\end{equation}
Here, the normalized firing strength serves as the weight, indicating the degree to which the antecedent conditions of rule $r_n$ are activated by each sample $X_m$. With decision scores, the consequent part of every rule can be reduced to a single representative value, improving the explainability of the IF-THEN rules and enabling their use in downstream LLMs.

Together, the state salience values and the IF-THEN rules with decision scores form the explanation interface of HFMAE-C. These structured explanations will be transformed by the LLM in User Interface described in the next section.

\section{LLM Agent-Supported User Interface}
The explanation interface of HFMAE-C provided the salience values and the IF-THEN rules. Both forms are understandable to human users, but their compact and highly structured representation makes them less intuitive, especially when a large number of states or rules are involved. To make these easier to use, and to exploit the unique advantage of the IF-THEN rules expressing in a natural language form that can be directly interpreted by LLMs, we build a user interface supported by multiple LLM-based agents. Through this interface, users can communicate with the system in natural language, describe the scale of explanation they need, and receive an explanation report written in an accessible form. Optional components further allow users to ask follow-up questions and obtain system diagnostics when needed.

\subsection{Overview}
\begin{figure}[htbp]
\centering
\includegraphics[width=0.5\textwidth]{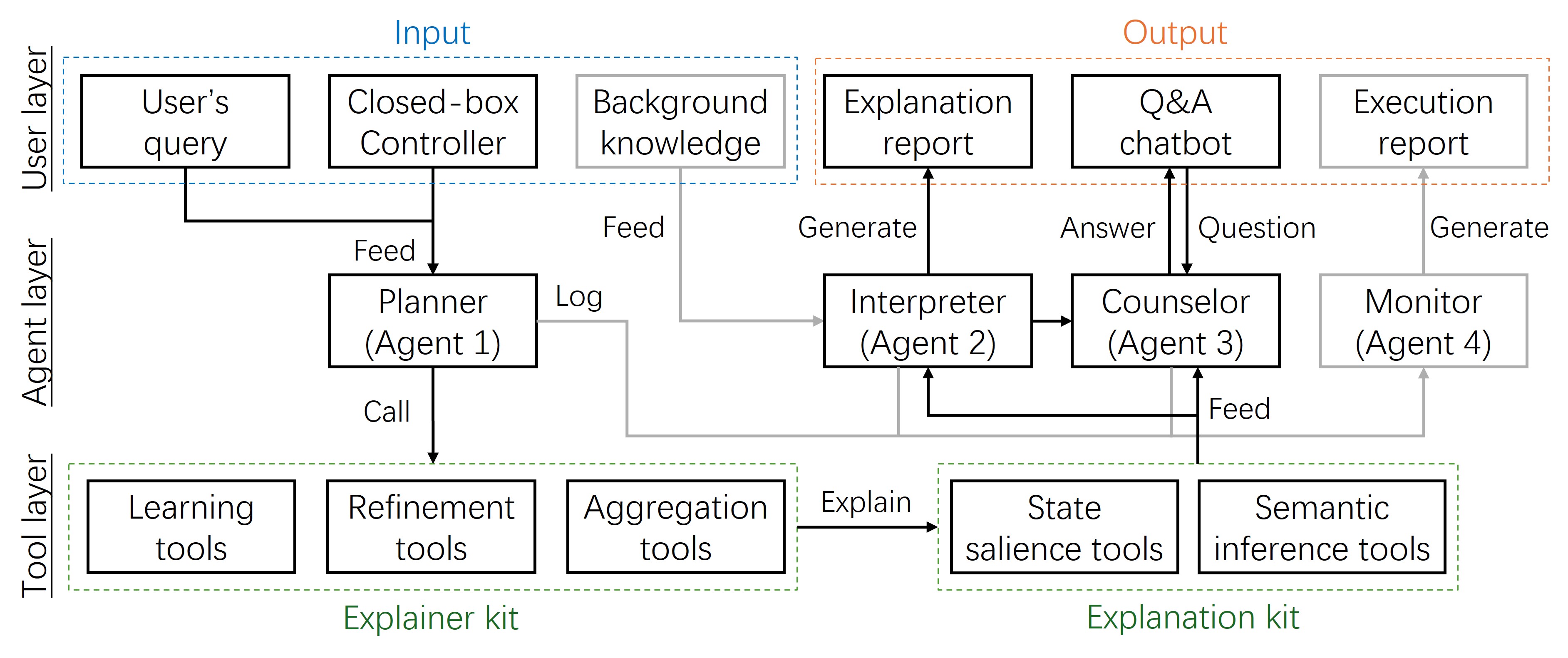}
\caption{Diagram of LLM agent-support user interface.}
\label{fig:ui}
\end{figure}

The user interface follows the three-layer structure shown in Fig. \ref{fig:ui}, where the components shown in black represent the core part of the system, while the components shown in grey are optional. The user layer includes what information the user needs to provide and what the system can return. The system accepts a natural-language query that states the user’s explanation requirement, a closed-box controller to be explained, and optional background knowledge that may help improve explanation quality. In return, the system produces a natural-language explanation report based on the explanation interface of HFMAE-C. When the optional agents are active, the user can also receive an interactive consultation and an execution report describing how the system operated.

The agent layer coordinates how the explanation is produced according with the user's requirement. The planner identifies the user’s intention, determines the explanation level, and selects the appropriate tools to construct the required explainer. The interpreter converts the structured outputs from HFMAE-C into a clear explanation report. Two additional agents serve as optional components: a counselor that answers user questions about the generated explanations, and a monitor that checks whether the system ran as expected and provides suggestions when the explanation quality needs improvement.

The tool layer corresponds directly to the HFMAE-C algorithm described earlier. It includes the explainer kit for training universe, domain, or local explainers through learning, refinement, and aggregation, and the explanation kit, which provides the salience values and IF-THEN rules produced by HFMAE-C. The agents interact with this layer to automatically call the tools consistent with user’s intention.

\subsection{Core Agents (Planner and Interpreter)}
The core of the proposed user interface is formed by two agents, the planner and the interpreter. They work together to transform a natural-language request into a trained explainer and, finally, into a readable explanation report.

\subsubsection{Planner}
The planner reads the user’s request and identifies the explanation scope required for the task. Its goal is to convert the request into a simple JSON object containing explanation scope, whose value must be universe, domain, or local. To guide the planner, a one-shot prompt is used. The prompt instructs the LLM to check whether the user explicitly mentioned the explanation scope and, if not, to infer the scope from the meaning of the request. An example of the prompt will be shown in Table I of the case study later. 
If a valid scope is obtained, the planner triggers the corresponding tool in the explainer kit to construct the explainer. If all attempts fail, the system assigns the default scope universe, since this level does not require the user to specify any instance or domain. The workflow is shown in Algorithm 1. 

\begin{algorithm}[!t]
\small
\caption{Planner Workflow}
\begin{algorithmic}[1]
    \Require User query $q$, maximum attempts $M$
    \Ensure Explanation scope $scope$ and trained explainer
    \State $attempts \gets 0$, $scope \gets$ None
    \While{$attempts < M$}
        \State Construct prompt using $q$
        \State $response \gets \text{LLM}(prompt)$
        \If{$response$ is a valid JSON object}
            \State $scope \gets response[\text{``scope''}]$
            \State \textbf{break}
        \EndIf
        \State $attempts \gets attempts + 1$
    \EndWhile
    \If{$scope$ is None}
        \State $scope \gets$ ``universe'' \Comment{fallback strategy}
    \EndIf
    \If{$scope$ = ``local''}
        \State Train local explainer by Learning and Refinement
    \ElsIf{$scope$ = ``domain''}
        \State Train domain explainer by Learning and Aggregation
    \Else
        \State Train universe explainer using Learning
    \EndIf
\end{algorithmic}
\label{algorithm_planner}
\end{algorithm}

\subsubsection{Interpreter}
Once the explainer is constructed, the interpreter generates the final explanation report. It receives the structured explanation interface of HFMAE-C, including state salience values and IF-THEN rules. 
The interpreter's prompt contains four element: optional background knowledge provided by the user, the formatted explanation results, a report template specifying the sections of the explanation document, and additional requirements such as writing style, tone, and length. A simple version of this prompt will be shown in Table I of case study later, and detailed prompt templates are provided in the \textit{Supplementary Materials}. 
The produced report is returned to the user as the main output of the interface.

\subsection{Optional Agents (Counselor and Monitor)}
In addition to the core agents, two optional agents may be activated to improve the usability and reliability of the system. These agents do not affect the main explanation pipeline but provide helpful support when users request additional clarification or the system needs to report its internal status.

The counselor is responsible for answering follow-up questions after an explanation report has been generated. It receives both the report and the user’s new question, and replies strictly based on the content of the explanation. This prevents the agent from introducing assumptions that are not supported by the explainer’s outputs. The counselor therefore acts as a natural-language interface that helps users understand specific parts of the explanation more deeply without requiring them to manually interpret the rule list or salience values.

The monitor focuses on system diagnostics. It reads the execution log produced during the explanation process, including information such as the explainer training steps, data usage, and tool outputs. When potential issues are detected, such as insufficient training samples or unusually high training error, the monitor generates a short execution report that highlights these signals and suggests possible remedies. This helps ensure that the explanation returned to the user is based on a well-functioning pipeline, and it provides transparency regarding how the system arrived at the final result.

\section{Experiments and Case Studies}

To comprehensively evaluate the proposed methods, experiments 
are conducted on two representative closed-loop control systems with different levels of complexity: a simple and intuitive control task is employed to demonstrate the hierarchical explanations provided by the proposed framework, while a more complex control problem is used to quantitatively evaluate explanation quality and compare with mainstream XAI methods. The following control systems are considered:

\begin{itemize}[leftmargin=*]

\item \textbf{Cart-pole inverted pendulum control:}
An inverted pendulum system is adopted from \cite{case}, whose control objective is to stabilize the pendulum around the upright position by applying a control force to the cart based on its angular displacement and angular velocity.

\item \Rf{\textbf{Turtlebot obstacle avoidance control using FAM-HGNN:}
A mobile robot navigation task is adopted from \cite{FAMHGNN}, where a Turtlebot robot navigates toward a target while avoiding three static obstacles. The controller is implemented using a fuzzy attention mechanism (FAM)-based heterogeneous graph neural network (HGNN) trained under, policy proximal optimization (PPO),  an RL algorithm. The control objective is to generate motion commands of robot wheels that enable collision-free navigation while reaching the target.}

\end{itemize}

The two control systems are selected for different experimental objectives. The inverted pendulum case exhibits intuitive behavior, making it suitable for demonstrating the hierarchical explanations generated by XCF. In contrast, the Turtlebot obstacle avoidance problem presents a considerably more challenging scenario for explainable control due to the uncertainty introduced by obstacle configurations and the complexity of the underlying controller, which combines RL with graph neural network-based relational reasoning. Therefore, this case is mainly employed to evaluate the explanation quality of XCF through quantitative comparison with representative XAI methods adopted by existing explainable control studies. Although both cases are used for explanation demonstration and quantitative evaluation, due to space limitations, representative explanation results are primarily presented for the inverted pendulum system, while quantitative evaluation results are emphasized for the Turtlebot case. The remaining results are provided in the \textit{Supplementary Materials}.

\subsection{Case Study I: Inverted Pendulum System}

\subsubsection{Control Task and Controller} 
This subsection presents a case study where the proposed framework is applied to a cart-pole type inverted pendulum system to explain the decision logic of its fuzzy controller and evaluate the performance of the trained explainers. The pendulum dynamics is described by two system state variables, namely the angular displacement $x_1$ and the angular velocity $x_2$. A control force $u$ is applied to the cart to stabilize the pendulum around $x_1=0$. Further details of the control system are provided in \cite{case}.

To demonstrate how the proposed framework can be used in practice, we design a simulated user experiment from the perspective of an engineer who aims to understand the decision logic of an opaque controller. His analysis follows a hierarchical process from global understanding to a specific scenario. The engineer first trains a universe explainer to obtain an overall view of the controller’s behavior. Based on this global understanding, the engineer then turns his attention to a specific operating scenario assumed to be relevant in his engineering application. To analyze the controller’s behavior in this region, multiple representative initial states are selected and explained using local explainers, which are further aggregated into a domain explainer that summarizes the controller’s behavior in this scenario. The results of the universe, local, and domain explainers are presented in the following parts.

\begin{figure}[!htb]
    \begin{minipage}[!htb]{\linewidth}
        \centering
        \includegraphics[width=0.5\textwidth]{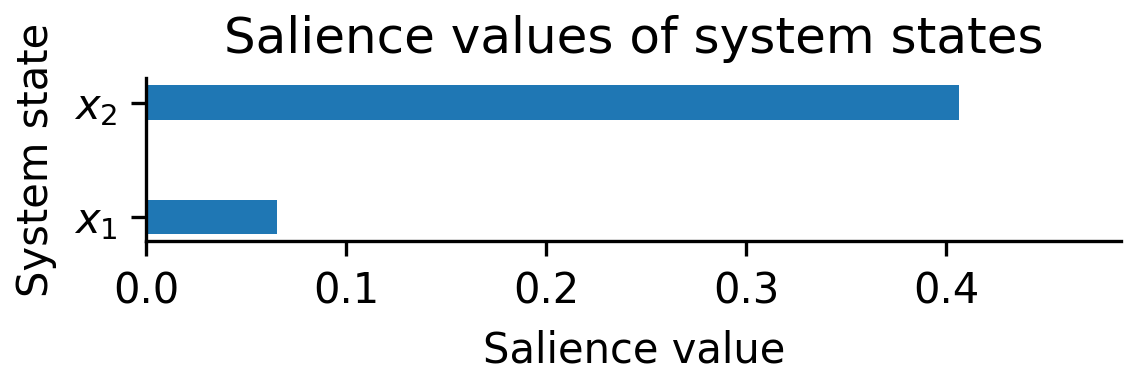}
        \centerline{(a) Salience values for universe level}
    \end{minipage}
    \begin{minipage}[!htb]{\linewidth}
        \centering
        \includegraphics[width=0.9\textwidth]{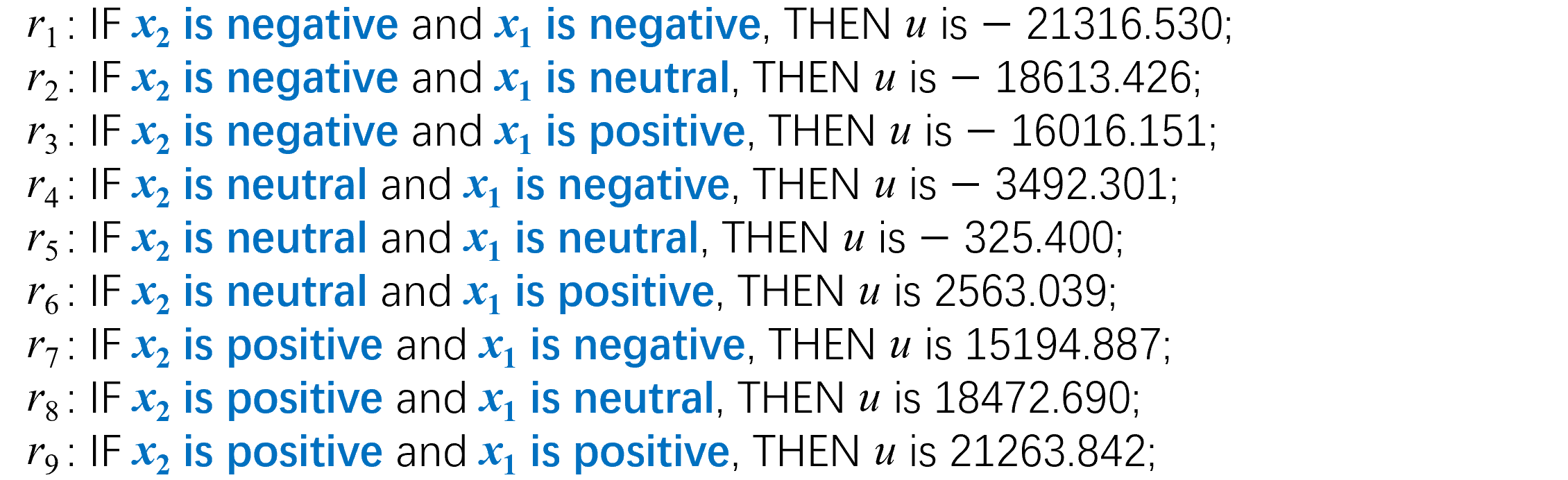}
        \centerline{(b) IF-THEN rules for universe level}
    \end{minipage}
    \begin{minipage}[!htb]{\linewidth}
        \centering
        \includegraphics[width=0.5\textwidth]{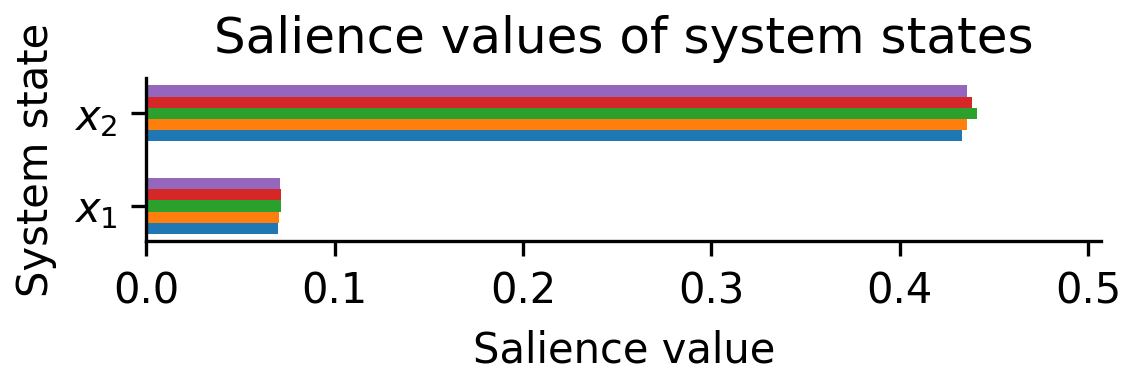}
        \centerline{(c) Salience values for local level}
    \end{minipage}
    \begin{minipage}[!htb]{\linewidth}
        \centering
        \includegraphics[width=0.9\textwidth]{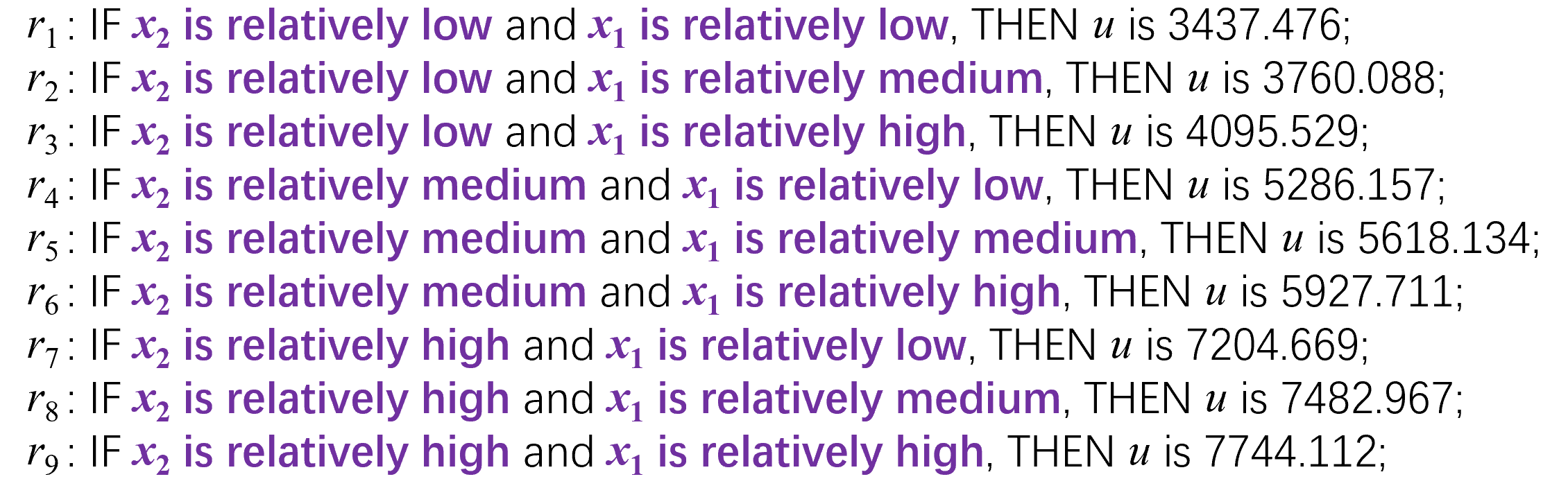}
        \centerline{(d) IF-THEN rules for local level}
    \end{minipage}
    \begin{minipage}[!htb]{\linewidth}
        \centering
        \includegraphics[width=0.5\textwidth]{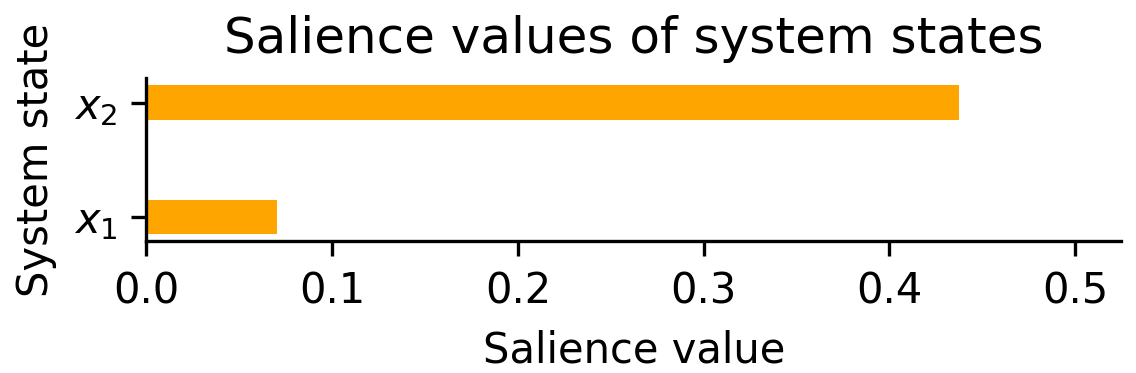}
        \centerline{(e) Salience values for domain level}
    \end{minipage}
    \begin{minipage}[!htb]{\linewidth}
        \centering
        \includegraphics[width=0.9\textwidth]{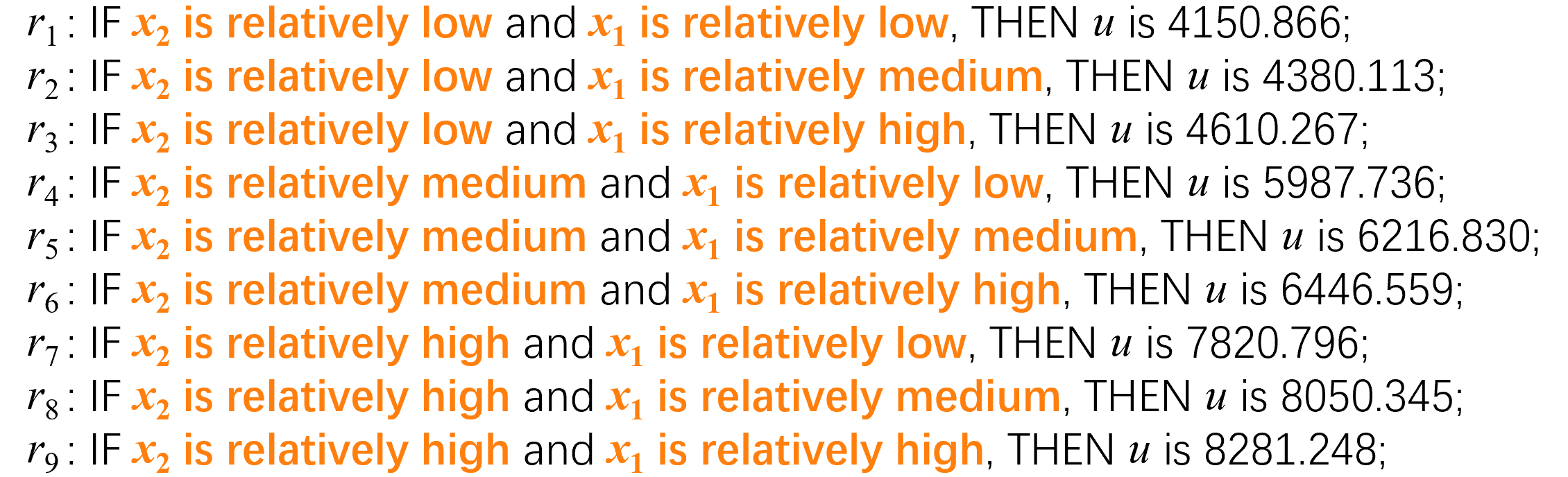}
        \centerline{(f) IF-THEN rules for domain level}
    \end{minipage}
    \caption{Hierarchical explanations for Case 1} 
    \label{fig:exp}
\end{figure}

\subsubsection{Universe Explainer}

The analysis begins with a universe explainer to obtain a global understanding of the controller’s decision logic across the entire state space. The state salience values shown in Fig. \ref{fig:exp}(a) indicate that both states positively contribute to the output, with the angular velocity \(x_2\) having higher salience than the angular displacement \(x_1\). This suggests that the controller primarily responds to the dynamic motion of the pendulum, while the angular displacement plays a secondary role. 
The semantic inference rules in Fig. \ref{fig:exp}(b) further reveal that the sign of the angular velocity largely determines the direction of the control force, while the angular displacement modulates its magnitude. To improve accessibility, the explanatory results can be further translated into concise natural language through the LLM-agent supported interface. Table \ref{tab:LLMs} presents a representative example generated by GPT-5.2 \cite{gpt52}. Additional examples and evaluation on multiple LLMs are reported in the \textit{Supplementary Materials}.

\subsubsection{Local Explainers}

Based on the global understanding, the analysis is further focused on a representative operating scenario characterized by a medium angular velocity and negative angular displacement, corresponding to a practical situation where the pendulum deviates from the upright position while continuing to move. Multiple local explainers are trained around representative states within this region to characterize the controller behavior in their neighborhoods. 
The resulting salience values are visualized in Fig. \ref{fig:exp}(c), showing consistent feature importance across different local explanations. A representative set of rules is shown in Fig. \ref{fig:exp}(d). Compared with universe explanation, these local rules provide more refined descriptions of how the controller adjusts its actions around specific operating conditions. LLM-assisted local explanations are provided in the \textit{Supplementary Materials}.

\begin{table}[h]
\centering
\caption{Prompts and Answers generated by LLMs}
\begin{tabular}{p{8cm}}
\hline
\textbf{User's Query}: Please explain the general behavior of the controller.
\\ \hline
\textbf{Planner Prompt}: Please extract the explanation scope from the user's query. 
The explanation scope must be one of: ``universe", ``domain", or ``local". 
First, you need to check whether the user has clearly mentioned the explanation scope (keywords: universe, domain, local). 
If not, you need to judge according to the following definition: 
the ``universe" scope is for global or general model behavior without specified instances, 
the ``domain" scope is for the model behavior on a group/set of instances, 
and the ``local" scope is for the model behavior on one instance. 
Respond in a JSON format like this:\\
\{``scope": ``domain"\} \\
Only respond with the JSON object.
\\ \hline
\textbf{Planner Answer}: \{``scope": ``universe"\}
\\ \hline
\textbf{Interpreter Prompt}: The following rules and salience values are the explanations of a controller behavior extracted by explainable control framework. Salience values represent the contribution of the features to the output. Based on these explanations, please summarize the behavior of controller to a paragraph with human-understandable form. \\ 
Background knowledge: (Provide background knowledge here)\\ 
Explanation results: (Provide explanations from HFMAE-C here)

\\ \hline
\textbf{Interpreter Answer}: 
The controller primarily bases its decisions on the pendulum’s angular velocity \(x_2\) , which has a much higher influence than the angular displacement 
\(x_1\). Overall, it applies forces to oppose the current motion of the pendulum and drive it back toward the upright position. When the pendulum is rotating in the negative direction, the controller consistently applies a strong negative force \(u\), with the magnitude increasing as the pendulum leans further to the negative side. When the angular velocity is positive, it applies a strong positive force \(u\), again scaling the force with the pendulum’s tilt. Near zero angular velocity, the controller uses much smaller corrective forces, pushing right or left depending on whether the pendulum leans negatively or positively. In essence, the controller behaves like a damping-dominated stabilizer that strongly counters pendulum motion while making fine position-based corrections near equilibrium.
\\ \hline
\end{tabular}
\label{tab:LLMs}
\end{table}

\subsubsection{Domain Explainer}

To summarize controller behavior within the selected operating scenario, the local explainers are aggregated into a domain explainer. The resulting salience values shown in Fig. \ref{fig:exp}(e) remain consistent with those observed at the local level, indicating that the aggregation preserves the dominant decision factors. The semantic inference rules in Fig. \ref{fig:exp}(f) provide a domain-level description of the controller’s decision logic, enabling reasoning about representative behavior within a region rather than around isolated states. Additional domain explanations are included in \textit{Supplementary Materials}.

\subsubsection{Performance Evaluation}

\begin{table}[h!]
\centering
\caption{Average Performance at Local Scale.}
\resizebox{0.8\columnwidth}{!}{
\begin{tabular}{c c c c c }
\hline
 \multirow{2}{*}{Explainer} 
    & \multicolumn{2}{c}{Prediction RMSE} 
    & \multicolumn{2}{c}{Response RMSE} \\
 & Training & Testing & $x_1$ & $x_2$ \\
\hline
 Universe & 102.6722 & 102.6438 & 0.0228 & 0.0095 \\
 Local (Step 1) & 5.0344 & 5.0047 & 0.0247 & 0.0095 \\
 Local (Step 2) & 8.2221 & 8.2170 & 0.0034 & 0.0018 \\
 Local (Step 3) & 2.5296 & 2.5351 & 0.0012 & 0.0010 \\
 Domain (Weight) & 11.8836 & 11.8822 & 0.0012 & 0.0010 \\
 Domain (Rule) & 1.5713 & 1.5604 & 0.0007 & 0.0007 \\
\hline
\end{tabular}}
\label{tab:result2}
\end{table}

Quantitative evaluation is conducted to assess how accurately the trained explainers reproduce the behavior of the original controller. Table \ref{tab:result2} reports the average prediction and response errors of explainers evaluated at 50 local-level instance (see detailed setting in \textit{Supplementary Materials}). 
The universe explainer achieves relatively larger prediction errors due to its broad approximation over the entire state space. In contrast, local explainers achieve substantially improved prediction accuracy by focusing on neighborhood behavior. The ablation results further demonstrate the contribution of each training step. While the local explainer trained only on local input-output data (Step 1) provides accurate local prediction, its response error remains relatively large. By incorporating global information during training (Step 2), response consistency is significantly improved. Finally, response-assisted refinement (Step 3) further reduces both prediction and response errors. 
The domain explainers preserve the advantages of local explainers while summarizing controller behavior within a broader operating region. In particular, rule aggregation achieves the lowest prediction and response errors, indicating that aggregating by fuzzy rules provides a more accurate approximation than weights. Complete evaluation results at universe and domain levels are provided in the \textit{Supplementary Materials}.

\begin{rf}
\subsection{Case Study II: Turtlebot Obstacle Avoidance}

\subsubsection{Control Task and Controller}

To evaluate the proposed method on a more complex control problem, we consider a Turtlebot obstacle avoidance navigation task based on the FAM-HGNN controller proposed in \cite{FAMHGNN}. 
The controller is trained using PPO, where the policy network is implemented HGNN. Specifically, the robot, target, and obstacles are represented as nodes in a fully connected heterogeneous graph, enabling the controller to capture their relational dependencies. FAM is employed to dynamically assign edge-wise fuzzy attention weights according to the relationship between entities, guiding message propagation within the HGNN.

The control objective is to navigate a differential-drive Turtlebot from a start location to a target location while avoiding collisions with surrounding obstacles. The environment consists of one robot, one target point, and three static obstacles in a 2-dimensional arena. The controller receives a 14-dimensional system state, 
including robot position and heading direction $(x_r,y_r)$ and $(h_x,h_y)$, left and right wheel velocities $(v_l,v_r)$, and the location of a target and 3 obstacles $(x_t,y_t)$ and $(x_{o_i},y_{o_i})$, where $i$ represents the $i$th obstacle; it outputs two wheel velocity commands for closed-loop navigation.

\subsubsection{Experimental Settings}

Since the original system state variables mainly describe physical configurations and do not directly convey semantically meaningful control intentions, the controller observations are first mapped into an explanatory feature space to describe controller's behavior. Based on the 14-dimensional state vector, five explanatory features are constructed to characterize the robot's navigation behavior, namely the distance and angular deviation to the target, the distance and angular deviation to the nearest obstacle, and the robot's linear speed, which are defined as follows:
\begin{equation}
\label{eq:turtlebot_feature_def}
\left\{
\begin{array}{l}
d_{\mathrm{target}}
=
\|\mathbf{p}_t-\mathbf{p}_r\|_2,\ 
\theta_{\mathrm{target}}
=
\angle(\mathbf{h},\mathbf{p}_t-\mathbf{p}_r),
\\
d_{\mathrm{obs}}
=
\min_i
\|\mathbf{p}_{o_i}-\mathbf{p}_r\|_2,\ 
\theta_{\mathrm{obs}}
=
\angle(\mathbf{h},\mathbf{p}_{o_j}-\mathbf{p}_r),
\\
v
=
\sqrt{v_x^2+v_y^2},
\end{array}
\right.
\end{equation}
where
$\mathbf{p}_r=(x_r,y_r)$,
$\mathbf{p}_t=(x_t,y_t)$,
and
$\mathbf{p}_{o_i}=(x_{o_i},y_{o_i})$
denote the robot, target, and obstacle positions, respectively,
$\mathbf{h}=[h_x,h_y]$
represents the robot heading direction,
$j=\arg\min_i \|\mathbf{p}_{o_i}-\mathbf{p}_r\|_2$
denotes the nearest obstacle,
and
$\angle(\cdot,\cdot)$
denotes the angle between two vectors. 
These features provide semantically meaningful descriptions of the robot's navigation state and are used as the explanation input space throughout the experiments. Since steering behavior is the primary decision of interest in obstacle avoidance navigation, the explanation target is defined as the absolute difference between the left and right wheel velocity commands $|u_l-u_r|$, where larger values indicate stronger turning actions.

To quantitatively evaluate explanation quality, experiments are conducted under an existing XAI evaluation framework \cite{metrics}. Within this framework, the proposed method is compared against XAI methods frequently adopted in explainable control studies, including SHAP \cite{SHAP}, LIME \cite{LIME}, and model agnostic supervised local explanations (MAPLE) \cite{MAPLE}. Two evaluation metrics are considered: \textit{Faithfulness} \cite{faithfulness}, which measures the Pearson correlation between feature importance and approximate marginal contribution, and \textit{Monotonicity} \cite{monotonicity}, which evaluates whether feature rankings correctly reflect their influence on controller output. Metric formulations and hyperparameter settings are reported in \textit{Supplementary Materials}.

\begin{figure}[!htb]
\centering
    \begin{minipage}[t]{0.45\linewidth}
        \centering
        \includegraphics[width=\textwidth]{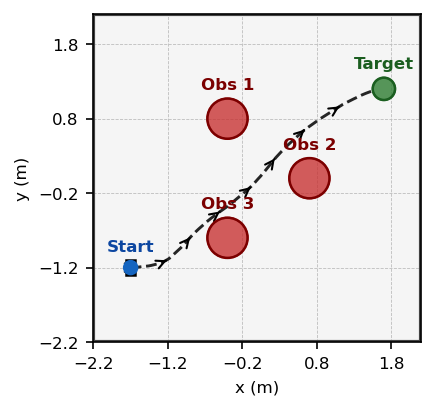}
        \centerline{\shortstack{(a) Map 1}}
    \end{minipage} 
    \begin{minipage}[t]{0.45\linewidth}
        \centering
        \includegraphics[width=\textwidth]{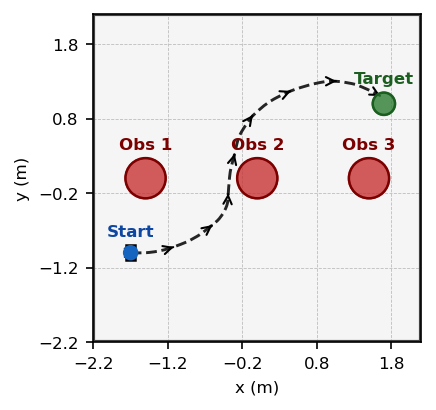}
        \centerline{\shortstack{(b) Map 2}}
    \end{minipage}
    \begin{minipage}[t]{0.45\linewidth}
        \centering
        \includegraphics[width=\textwidth]{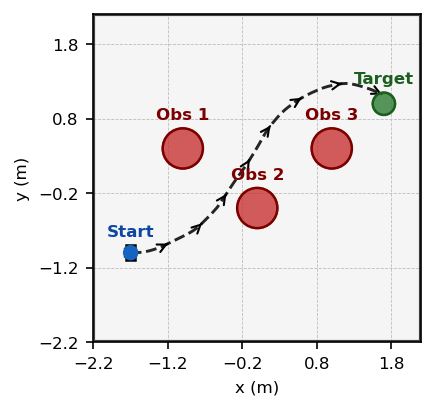}
        \centerline{\shortstack{(c) Map 3}}
    \end{minipage} 
    \begin{minipage}[t]{0.45\linewidth}
        \centering
        \includegraphics[width=\textwidth]{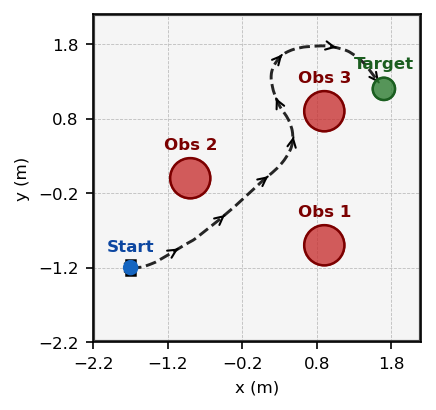}
        \centerline{\shortstack{(d) Map 4}}
    \end{minipage}   
    \caption{Experimental environments with 4 obstacle layouts.}
    \label{fig:map}
\end{figure}

\begin{table}[h]
\centering
\caption{XAI Methods Comparison on Turtlebot Control}
\label{tab:xai_comparison}
\resizebox{0.8\columnwidth}{!}{
\begin{tabular}{llccc}
\toprule
Case & Method & \multicolumn{2}{c}{State Salience} & Approximation \\ 
 &  & Faithfulness & Monotonicity & $R^2$ Score \\ 
\midrule
\multirow{4}{*}{Map 1} & SHAP & 0.540$\pm$0.170 & 0.600$\pm$0.063 & \textemdash \\ 
 & LIME & 0.452$\pm$0.231 & 0.584$\pm$0.037 & 0.235$\pm$0.045 \\ 
 & MAPLE & 0.369$\pm$0.207 & 0.604$\pm$\textbf{0.027} & \textemdash \\ 
 & XCF & \textbf{0.615}$\pm$\textbf{0.109} & \textbf{0.618}$\pm$0.051 & \textbf{0.743}$\pm$\textbf{0.024} \\ 
\midrule
\multirow{4}{*}{Map 2} & SHAP & 0.578$\pm$0.115 & 0.595$\pm$0.053 & \textemdash \\ 
 & LIME & 0.519$\pm$\textbf{0.080} & 0.603$\pm$0.050 & 0.294$\pm$0.045 \\ 
 & MAPLE & 0.444$\pm$0.164 & 0.608$\pm$0.042 & \textemdash \\ 
 & XCF & \textbf{0.665}$\pm$0.093 & \textbf{0.624}$\pm$\textbf{0.035} & \textbf{0.802}$\pm$\textbf{0.029} \\ 
\midrule
\multirow{4}{*}{Map 3} & SHAP & 0.590$\pm$\textbf{0.147} & 0.586$\pm$0.083 & \textemdash \\ 
 & LIME & 0.478$\pm$0.163 & 0.566$\pm$0.050 & 0.277$\pm$0.057 \\ 
 & MAPLE & 0.443$\pm$0.210 & 0.602$\pm$0.037 & \textemdash \\ 
 & XCF & \textbf{0.609}$\pm$0.162 & \textbf{0.605}$\pm$\textbf{0.022} & \textbf{0.786}$\pm$\textbf{0.028} \\ 
\midrule
\multirow{4}{*}{Map 4} & SHAP & 0.543$\pm$0.174 & 0.597$\pm$0.051 & \textemdash \\ 
 & LIME & 0.470$\pm$0.203 & 0.588$\pm$\textbf{0.051} & 0.280$\pm$0.071 \\ 
 & MAPLE & 0.394$\pm$0.227 & 0.584$\pm$0.065 & \textemdash \\ 
 & XCF & \textbf{0.658}$\pm$\textbf{0.073} & \textbf{0.606}$\pm$0.077 & \textbf{0.753}$\pm$\textbf{0.032} \\ 
\bottomrule
\end{tabular}}
\end{table}

\subsubsection{Performance Evaluation}

The experimental environments considered in this study are illustrated in Fig. \ref{fig:map}, with four representative obstacle layouts. For each map, 20 positions are sampled around the start point, target point, and each of the three obstacles, resulting in 100 local regions for controller explanation. Local explanations generated by different XAI methods are evaluated and the results are summarized in Table \ref{tab:xai_comparison}. Since both LIME and the proposed method rely on surrogate models for explanation, their approximation performance is additionally compared using the surrogate prediction \(R^2\) score. For each metric, both the mean and standard deviation across the 100 local explanations are reported.

From Table \ref{tab:xai_comparison}, it can be observed that the proposed XCF method consistently achieves the best average performance across all four maps and all evaluation metrics. XCF obtains the highest \textit{Faithfulness} in every case, indicating that the generated feature salience values more accurately reflect the controller's actual decision basis. Similarly, XCF achieves the highest average \textit{Monotonicity} scores across all obstacle layouts, suggesting that the ranking of important features better matches their true influence on control outputs. Although several baseline methods occasionally exhibit slightly lower standard deviations in specific cases, XCF maintains competitive or lower variance in most settings, demonstrating stable explanation quality under different configurations. Furthermore, compared with the linear surrogate adopted by LIME, the employed fuzzy surrogate achieves higher approximation accuracy across all maps, showing considerably more capable of reconstructing the non-linear control behavior. 

Overall, the experimental results demonstrate that XCF can generate more faithful and reliable explanations for complex control systems. Due to space limitations, simulated user experiment with comprehensive explanations generated by XCF are provided in the \textit{Supplementary Materials}.
\end{rf}

\section{Conclusion and Discussion}
This paper is an extension of a recent conference publication \cite{xcf_c}. We propose the XCF that systematically extends model-agnostic XAI techniques to closed-loop control systems, aiming to address the challenges of universality and accessibility in explainable control. By treating the controller as a closed box, XCF enables hierarchical explanations of control behavior without requiring access to internal structures. As a concrete instantiation of the framework, the HFMAE-C is developed, using FLSs to approximate controller behavior and provide sample, local, domain, and universe-level explanations through IF–THEN rules and state salience values. Furthermore, an LLM agent-supported user interface is introduced to bridge the gap between technical explanations and user understanding, allowing non-expert users to specify explanation requirements and receive structured, accessible reports and interactive consultation. \Rf{Two case studies demonstrated that the proposed framework achieves high approximation accuracy while offering meaningful and consistent explanations across scales.} Future work will explore deeper integration between LLMs and explainable control, such as using LLMs for higher-level semantic abstraction and user-machine interaction, allowing users to reason about similarities and differences in controller behavior across operating domains.

\section*{Acknowledgment}
Faliang Yin contributed to algorithm details and refinement, experiments and led paper writing. Hak-Keung Lam conceived, proposed and developed the FMAE and XCF frameworks, and algorithms, and paper refinement. David Watson provided feedback on the methods and assisted in writing.

\ifCLASSOPTIONcaptionsoff
  \newpage
\fi

\bibliographystyle{ieeetr}  
\normalem
\bibliography{ref}

\end{document}


\renewcommand{\thesection}{S.\Roman{section}} 
\renewcommand{\thesubsection}{\thesection.\Alph{subsection}}

\makeatletter
\renewcommand{\thetable}{S.\Roman{table}}
\renewcommand{\theequation}{S\arabic{equation}}
\makeatother

\title{Supplementary Materials of `Explainable Control Framework (XCF) based on Fuzzy Model-Agnostic Explanation and LLM Agent-Supported Interface'}

\author{Faliang Yin, Hak-Keung Lam,~\IEEEmembership{Fellow,~IEEE}, David Watson
\vspace{-1em}
\thanks{This research is partially supported by King’s-China Scholarship Council PhD Scholarship programme (K-CSC). Acknowledgement: Faliang Yin contributed to algorithm details and refinement, experiments and led paper writing. Hak-Keung Lam conceived, proposed and developed the FMAE and XCF frameworks, and algorithms, and paper refinement. David Watson provided feedback on the methods and assisted in the writing process. \textit{(Corresponding author: Hak-Keung Lam.)}}
\thanks{Faliang Yin and  Hak-Keung Lam are with the Department of Engineering,
Faculty of Natural, Mathematical \& Engineering Sciences, King's College London, WC2R 2LS London, United Kingdom (e-mail: faliang.yin@kcl.ac.uk; hak-keung.lam@kcl.ac.uk).}
\thanks{David Watson is with the Department of Informatics, Faculty of Natural, Mathematical \& Engineering Sciences, King's College London, WC2R 2LS London, United Kingdom (e-mail: david.watson@kcl.ac.uk).}
}

\markboth{Preprint submitted to IEEE for possible publication.}%
{Shell \MakeLowercase{\textit{et al.}}: Bare Demo of IEEEtran.cls for IEEE Journals}

\maketitle

\begin{center} \begin{minipage}{0.95\linewidth} \small \textbf{Preprint Notice} This work has been submitted to the IEEE for possible publication. Copyright may be transferred without notice, after which this version may no longer be accessible. \end{minipage} \end{center}

This supplementary material provides additional experimental details and results that complement the main paper and are omitted from the main text due to space limitations. Section S.I evaluates the performance of different large language models (LLMs) used as the planner in the user interface, focusing on feasibility and reliability under limited computational resources. Section S.II extends Case Study I on the inverted pendulum system by providing supplementary experimental settings, additional simulated user experiments, and detailed performance evaluation results omitted from the main text. \Rf{Section S.III extends Case Study II on Turtlebot obstacle avoidance by presenting supplementary experimental settings and additional simulated user experiments, including explanations for representative obstacle avoidance scenarios.}

\begin{figure*}[htbp]
\centering
\includegraphics[width=1\textwidth]{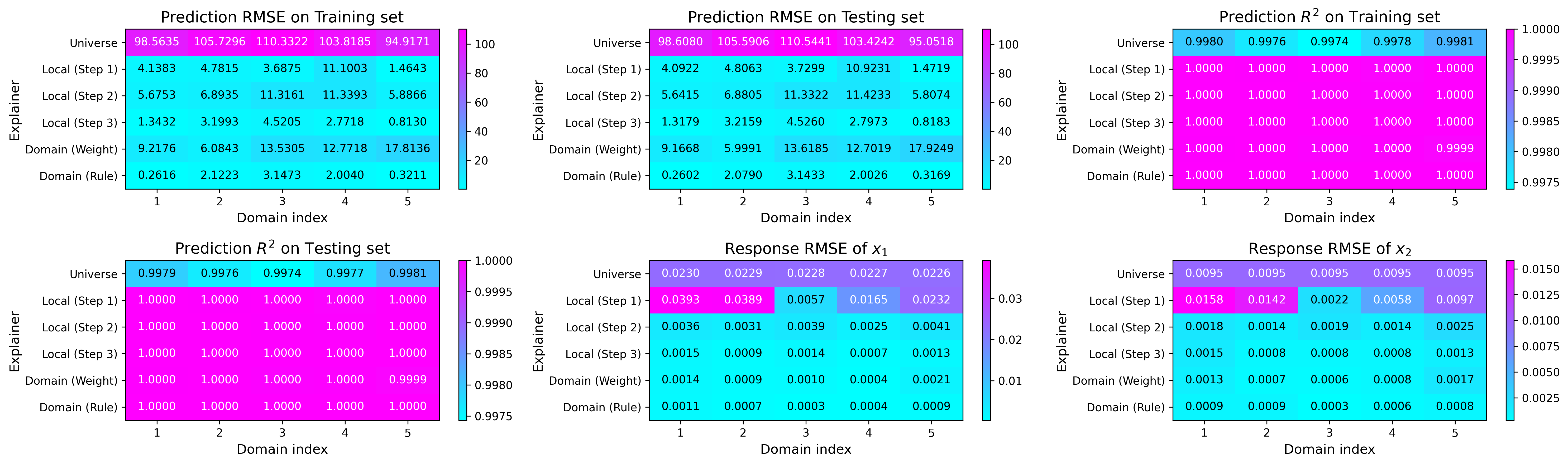}
\caption{Average Performance at Local Scale per Domain.}
\label{fig:pfm}
\end{figure*}

\section{Evaluation of LLMs Used as the Planner}

This section supplements the evaluation of the large language model based planner introduced in Section IV and partially validated in Section V.A of the main paper. While the main paper demonstrates the feasibility of the planner using a flagship LLM (GPT-5.2 \cite{gpt52}), the focus here is on assessing its deployability under limited computational resources. Specifically, we investigate whether lightweight LLMs with small memory footprints can reliably perform the planner’s function, which is essential for deployment on resource-constrained platforms.

From a functional perspective, the planner addresses an intent recognition problem, where the goal is to infer the intended explanation scope, namely universe, domain, or local, from a natural-language user request. To evaluate this capability, a test set of 1000 natural-language queries is randomly generated with ground-truth labels indicating the correct explanation level. The requests cover representative user intents at all three levels, for example, queries asking for explanations of the controller behavior at a specific instance or the controller's general decision logic.

Multiple LLMs with parameter sizes no larger than 4 billion (4b) are selected as planner candidates, all of which require relatively low GPU memory and are suitable for deployment on lightweight devices. Each LLM is used to infer the explanation scope and generate a JSON object for tool invocation.

Two metrics are adopted for evaluation. The \emph{Success rate} measures the reliability of tool invocation and is defined as
\begin{equation}
\mathrm{Success\ rate}=\frac{N_{\mathrm{valid}}}{N_{\mathrm{total}}},
\end{equation}
where $N_{\mathrm{valid}}$ is the number of requests for which the LLM produces a syntactically valid and executable JSON output, and $N_{\mathrm{total}}=1000$ is the total number of test requests.

The \emph{Accuracy} measures the correctness of intent recognition and is defined as
\begin{equation}
\mathrm{Accuracy}=\frac{N_{\mathrm{correct}}}{N_{\mathrm{total}}},
\end{equation}
where $N_{\mathrm{correct}}$ denotes the number of requests for which the inferred explanation level matches the annotated ground-truth label. This metric is reported both overall and separately for each explanation level.

The evaluation results are summarized in Table~\ref{tab:LLMs}, where the accuracy is reported separately for the local, domain, and universe levels in different columns, and the overall performance is summarized in the \emph{Total} column. All tested models achieve near-perfect success rates, indicating that lightweight LLMs can robustly generate tool-invocable JSON outputs. In terms of accuracy, all models correctly identify universe- and local-level intents with 100\% accuracy. In contrast, misclassifications are observed primarily at the domain level for some models. This is mainly because the notion of a domain is relatively more complex and may introduce to ambiguity in natural-language expressions. In particular, when a user provides a set of instances without explicitly stating the need for a domain explanation, for example, describing the request using terms such as a region, some LLMs may misinterpret it as a collection of local-level queries. Nevertheless, for most evaluated models, the achieved domain-level accuracy still indicates that the planner remains practically feasible under limited computational resources. In future work, it could be considered integrating existing intent recognition enhancement techniques to further improve domain-level accuracy and robustness across models.

\begin{table}[ht]
\centering
\caption{Planner performance using different LLMs}
\resizebox{\columnwidth}{!}{
\begin{tabular}{cccccc}
\hline
\multirow{2}{*}{Model} & \multirow{2}{*}{Success   rate} & \multicolumn{4}{c}{Accuracy}              \\
                       &                                 & Local    & Domain   & Universe & Total    \\ \hline
Qwen3   (4b) \cite{Qwen3}    & 100.00\%                        & 100.00\% & 100.00\% & 100.00\% & 100.00\% \\
Gemma3n   (4b) \cite{gemma3}         & 100.00\%                        & 100.00\% & 100.00\% & 100.00\% & 100.00\% \\
Gemma3   (4b) \cite{gemma3}          & 100.00\%                        & 100.00\% & 99.40\%  & 100.00\% & 99.80\%  \\
Llama3.2   (3b) \cite{llama3}        & 100.00\%                        & 100.00\% & 91.90\%  & 100.00\% & 97.20\%  \\
Phi4-mini   (3.8b) \cite{phi4mini}     & 99.80\%                         & 100.00\% & 84.00\%  & 100.00\% & 94.50\%  \\ \hline
\end{tabular}}
\label{tab:LLMs}
\end{table}

\section{Extension of Case Study I: Inverted Pendulum System}
\begin{figure}[htbp]
\centering
\includegraphics[width=0.25\textwidth]{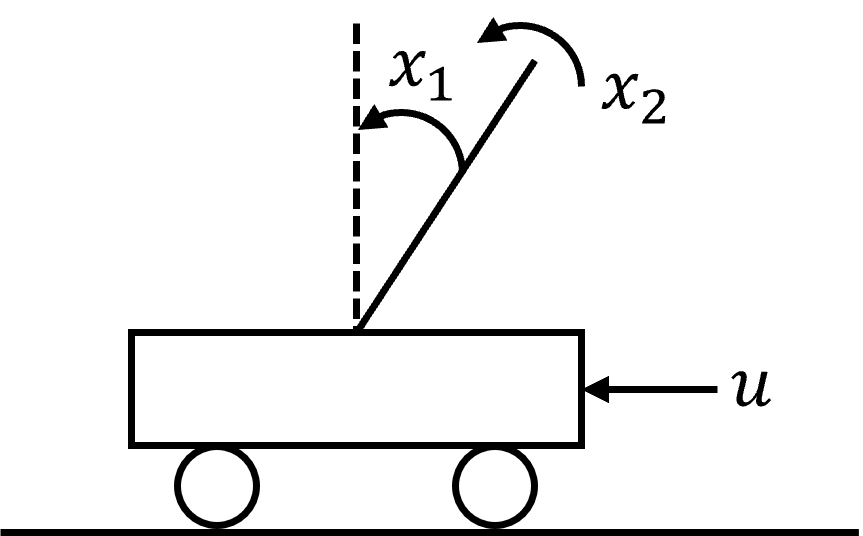}
\caption{Cart-pole type inverted pendulum system.}
\label{fig:cart}
\end{figure}

This section provides experimental settings and additional experimental results that extend the case study presented in Section V.A of the main paper. All results reported here serve as supplementary evidence to the simulated user experiment and the performance evaluation, and support the same conclusions drawn in the main text.

\subsection{Experimental Settings}
\subsubsection{Hyperparameter Settings} This part supplements the hyperparameter settings used in the case study and performance evaluation reported in Section V.A of the main paper. Table \ref{tab:para} summarizes the dataset configuration and the hyperparameter settings adopted for the explainers and the associated optimization procedures. For each explanation level, namely universe, domain, and local, 5000 samples are collected and randomly split into training and testing sets with a ratio of 4:1. The remaining parameters specify the structure of the fuzzy logic system (FLS) and the optimization settings for universe-scale learning, local-scale learning, and the subsequent refinement steps, and are kept consistent across all experiments.
\begin{table}[ht]
\centering
\caption{Hyperparameter Settings}
\begin{tabular}{cc}
\hline
Hyperparameter                          & Value \\ \hline
 Dataset size for local, domain and universe level&5000\\
 Training/Testing Split&4:1\\ \hline
 Number of fuzzy sets for each state&3\\
 Maximum iterations for universe-scale learning   & 2000\\
Maximum iterations for local-scale learning   & 2000\\
Maximum iterations for global-assisted refinement& 500  
\\
 Maximum iterations for response-assisted refinement&50\\
 Learning rate for universe-scale learning &0.2\\
 Learning rate for local-scale learning &0.2\\
 Learning rate for refinement steps&0.001\\  \hline
\end{tabular}
\label{tab:para}
\end{table}

\subsubsection{Formulation of FLS} 
For completeness, we provide the full formulation of the FLS used as the explainer in HFMAE-C. The FLS $\hat{u}(\mathbf{x})$ employed as the explainer is formed by a fully combined fuzzy rule base $R=\{r_1,r_2,...,r_N\}$, and the overall output of  $\hat{u}(\mathbf{x})$ is the ensemble of all the $N$ fuzzy rules. The typical linguistic form of a TSK \cite{TSK} fuzzy rule $r_n$ consists of an antecedent (IF part) and a consequent (THEN part):
\begin{align}
r_n: ~&IF\ x_1\ is\ A^1_n\ and\ ...\ and\ x_D\ is\ A^D_n,\nonumber \\
&THEN\ f_{n}(\mathbf{x}) =  a^0_{n}+a_{n}^1x_1+a_{n}^2x_2+...+a_{n}^Dx_D. 
\label{eq:fuzzymodel_rule_n}
\end{align}
In the antecedent, $x_d$ is the $d$th system state, and $A^d_n$ is the fuzzy set of $d$th system state in the $n$th fuzzy rule, where $d=1,2,...,D$ and $n=1,2,...,N$. For each premise in the antecedent like ``$x_d\ is\ A^d_n$'', its degree of truth, denoted as $\mu^d_n(x_d)$, is measured by the membership function (MF), which contributes to the robustness by flexibly modeling uncertainty between system states and linguistic terms. There are various choices to calculate the MF value, and here we use a simplified Gaussian MF for example. The MF value of $x_d$ to the fuzzy set $A^d_n$ with a center $c^d_n$ can be calculated by
\begin{equation}
\mu^d_n(x_d) = e^{-(x_d-c^d_n)^2},
\label{MF}
\end{equation}
In the consequent, $f_{n}(\mathbf{x})$ is a trainable linear model in TSK fuzzy rules, where the consequent parameters are $a^d_{n}$ denoting a scalar coefficient for the $d$th system state, and $a^0_{n}$ denoting a scalar bias. The $and$ operator combines all the premises by  t-norm \cite{TSK} operation like the product of their MF values as joint effect:
\begin{equation}
\mu_n(\mathbf{x})=\prod_{d=1}^D\mu^d_n(x_d),\ \mu_n(\mathbf{x})\in [0, 1].
\label{miu}
\end{equation}
$\mu_n(\mathbf{x})$ is known as firing strength, and after normalization it can represent the activation level of the $n$th rule:
\begin{equation}
\omega_n(\mathbf{\mathbf{x}})=\frac{\mu_n(\mathbf{x})}{\sum^N_{n=1}\mu_n(\mathbf{x})},\ \omega_n(\mathbf{x})\in[0,1].
\label{fire}
\end{equation}
The output is the weighted average output of all the $N$ rules:
\begin{equation}
\hat{u}(\mathbf{x})=\sum^N_{n=1}\omega_n(\mathbf{x}) ~f_{n}(\mathbf{x}).
\label{out}
\end{equation}

This formulation is consistent with the explainer used in the main text. In HFMAE-C, the FLS serves as a surrogate model for approximating the behavior of the target controller, and all hierarchical explanation levels are constructed based on this unified representation.

\subsection{Extension of Simulated User Experiment}

This part supplements the simulated user experiment reported in Section V.A of the main paper. In Table \ref{tab:local}, we present an example of the prompt and the corresponding output used to generate a local-level explanation report for the controller. The generated report consists of multiple paragraphs and provides a more detailed description of the controller’s decision rationale and underlying logic at a specific instance. Compared with the single-paragraph universe-level report shown in the main paper, the local-level report offers a finer-grained and more focused explanation. The prompt is constructed using the markdown format commonly adopted in LLM-based applications, which facilitates structured outputs.

Tables \ref{tab:domain_pmt} and \ref{tab:domain_ans} further extend the results by illustrating the prompt and the corresponding output used to generate domain-level reports. In this setting, a report template is explicitly provided in the prompt to guide the LLM to produce a structured explanation. This design enables users to obtain a detailed and organized report that focuses on the controller’s behavior within a specific domain of interest. The presented results are consistent with the domain explanations discussed in the main paper and further demonstrate the flexibility of the proposed interface in supporting user's requirement.

\subsection{Extension of Performance Evaluation}

\begin{table*}[h!]
\centering
\caption{Average Performance at Different Scales.}
\begin{tabular}{c c c c c c c c}
\hline
\multirow{2}{*}{Scale} 
& \multirow{2}{*}{Explainer} 
& \multicolumn{2}{c}{Prediction RMSE} 
& \multicolumn{2}{c}{Prediction $R^2$ score} & \multicolumn{2}{c}{Response RMSE}\\
&  & Training & Testing & Training & Testing  & $x_1$&$x_2$\\
\hline
Universe 
& Universe& 161.9010 & 157.9501 & 1.0000 & 1.0000  & --&--\\
\hline
\multirow{2}{*}{Domain}
& Domain (Weight)& 12.5214 & 12.7981 & 1.0000 & 1.0000  & --&--\\
& Domain (Rule)& 1.6143 & 1.6799 & 1.0000 & 1.0000  & --&--\\
\hline
 \multirow{6}{*}{Local}&Universe & 102.6722 & 102.6438 & 0.9978 & 0.9978 & 0.0228 & 0.0095 \\
 &Local (Step 1) & 5.0344 & 5.0047 & 1.0000 & 1.0000 & 0.0247 & 0.0095 \\
 &Local (Step 2) & 8.2221 & 8.2170 & 1.0000 & 1.0000 & 0.0034 & 0.0018 \\
 &Local (Step 3) & 2.5296 & 2.5351 & 1.0000 & 1.0000 & 0.0012 & 0.0010 \\
 &Domain (Weight) & 11.8836 & 11.8822 & 1.0000 & 1.0000 & 0.0012 & 0.0010 \\
 &Domain (Rule) & 1.5713 & 1.5604 & 1.0000 & 1.0000 & 0.0007 & 0.0007 \\
\hline
\end{tabular}
\label{tab:result1}
\end{table*}

This part supplements the performance evaluation reported in Section V.A in the main paper. While the main paper focuses on the averaged local-scale performance of the explainers, this section further presents detailed quantitative results at the universe and domain levels, together with additional local-scale analyses omitted from the main text due to space limitations.

The performance evaluation is conducted by comparing the outputs of the trained explainers with those of the original controller. First, a universe explainer is trained and tested on data sampled from the full state space. Next, 50 representative initial states are sampled to construct local-scale test cases, as listed in Table \ref{tab:cases}, where each initial state is used to train a corresponding local explainer. Based on the identified importance of the angular velocity $x_2$ from the universe explanation, the 50 local cases are grouped according to the value of $x_2$, with every ten cases forming one domain, representing a distinct operating region in the state space. For each domain, the corresponding local explainers are aggregated to construct two domain explainers using weight aggregation and rule aggregation, respectively. The quantitative performance results at different explanation levels are summarized in Table~\ref{tab:result1}, while detailed prediction performance grouped by domain is further visualized in Fig.~\ref{fig:pfm}.

\begin{table}[!htb]
\centering
\caption{Initial Conditions for Simulation.}
\resizebox{\columnwidth}{!}{
\begin{tabular}{cccccc}
\hline
 & Doamin 1 & Doamin 2 & Doamin 3 & Doamin 4 & Doamin 5 \\  \hline
0 & ($-$10$\pi$/45, $-$2.0) & ($-$10$\pi$/45, $-$1.0) & ($-$10$\pi$/45, 0) & ($-$10$\pi$/45, 1.0) & ($-$10$\pi$/45, 2.0) \\
1 & ($-$8$\pi$/45, $-$2.0) & ($-$8$\pi$/45, $-$1.0) & ($-$8$\pi$/45, 0) & ($-$8$\pi$/45, 1.0) & ($-$8$\pi$/45, 2.0) \\
2 & ($-$6$\pi$/45, $-$2.0) & ($-$6$\pi$/45, $-$1.0) & ($-$6$\pi$/45, 0) & ($-$6$\pi$/45, 1.0) & ($-$6$\pi$/45, 2.0) \\
3 & ($-$4$\pi$/45, $-$2.0) & ($-$4$\pi$/45, $-$1.0) & ($-$4$\pi$/45, 0) & ($-$4$\pi$/45, 1.0) & ($-$4$\pi$/45, 2.0) \\
4 & ($-$2$\pi$/45, $-$2.0) & ($-$2$\pi$/45, $-$1.0) & ($-$2$\pi$/45, 0) & ($-$2$\pi$/45, 1.0) & ($-$2$\pi$/45, 2.0) \\
5 & (2$\pi$/45, $-$2.0) & (2$\pi$/45, $-$1.0) & (2$\pi$/45, 0) & (2$\pi$/45, 1.0) & (2$\pi$/45, 2.0) \\
6 & (4$\pi$/45, $-$2.0) & (4$\pi$/45, $-$1.0) & (4$\pi$/45, 0) & (4$\pi$/45, 1.0) & (4$\pi$/45, 2.0) \\
7 & (6$\pi$/45, $-$2.0) & (6$\pi$/45, $-$1.0) & (6$\pi$/45, 0) & (6$\pi$/45, 1.0) & (6$\pi$/45, 2.0) \\
8 & (8$\pi$/45, $-$2.0) & (8$\pi$/45, $-$1.0) & (8$\pi$/45, 0) & (8$\pi$/45, 1.0) & (8$\pi$/45, 2.0) \\
9 & (10$\pi$/45, $-$2.0) & (10$\pi$/45, $-$1.0) & (10$\pi$/45, 0) & (10$\pi$/45, 1.0) & (10$\pi$/45, 2.0) \\ \hline
\end{tabular}
}
\label{tab:cases}
\end{table}

\subsubsection{Prediction error}

Prediction error is first evaluated to assess how accurately each explainer approximates the control signal produced by the original controller. The evaluation is conducted at three different scales, namely universe, domain, and local, where each explainer is tested at the scale for which it is designed, as well as at the local scale where a unified comparison becomes possible.

At the universe and domain scales, the prediction performance is summarized in Table~\ref{tab:result1}. The universe explainer achieves an $R^2$ score close to 1 on both the training and testing sets, indicating that it captures the global input--output relationship of the controller over the entire state space with high fidelity. Although the prediction error measured by root mean square error (RMSE) is relatively large in magnitude, this is expected because the control signal spans a wide numerical range at the global scale. Nevertheless, the strong predictive performance suggests that the universe explainer can still effectively approximate the controller’s overall decision logic while maintaining interpretability.

Compared with the universe explainer, the domain explainers consistently achieve substantially lower prediction RMSE while maintaining near-perfect $R^2$ scores, indicating that restricting the explanation scope to a representative operating region significantly improves approximation accuracy. This observation is consistent with the hierarchical design philosophy of the proposed framework, namely that narrowing the explanation scope enables the explainer to better capture localized controller behavior. Furthermore, among the two aggregation strategies, rule aggregation consistently achieves lower RMSE than weight aggregation, suggesting that aggregating semantic rules preserves controller behavior more effectively than directly averaging model parameters.

At the local scale, the main paper reports the averaged prediction performance over all 50 local instances. To provide a more detailed view, Fig.~\ref{fig:pfm} presents the prediction error results grouped by domain. The figure is organized as a grid with six subplots, each corresponding to a different performance metric. As observed from the figure, the predictive performance remains highly consistent across different operating regions, following trends similar to the averaged results reported in the main text. In particular, local explainers after refinement consistently achieve high predictive accuracy across all domains, indicating that the proposed training strategy maintains stable approximation performance under different operating conditions.

In addition, the domain explainers obtained via weight aggregation exhibit intermediate performance between universe and local explainers, indicating that simple parameter averaging partially preserves local information but does not fully exploit the underlying controller structure. By contrast, the domain explainer obtained through rule aggregation consistently achieves superior approximation performance. This result further highlights the advantage of the proposed rule aggregation strategy, which enables the explainer to capture the semantic relationship among local control behaviors and construct a more accurate domain-level approximation.

\subsubsection{Response error}

\begin{figure*}[!htb]
    \begin{minipage}[!htb]{\linewidth}
        \centering
        \includegraphics[width=0.95\textwidth]{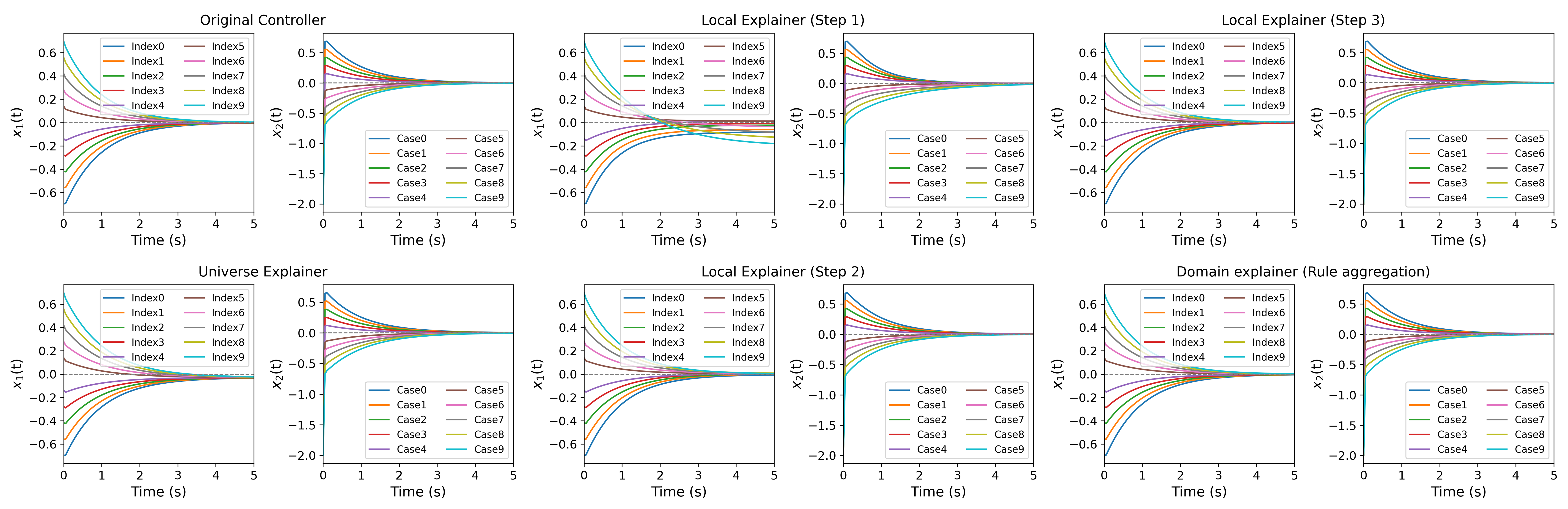}
        \centerline{(a) Domain 1}
    \end{minipage}
    \begin{minipage}[!htb]{\linewidth}
        \centering
        \includegraphics[width=0.95\textwidth]{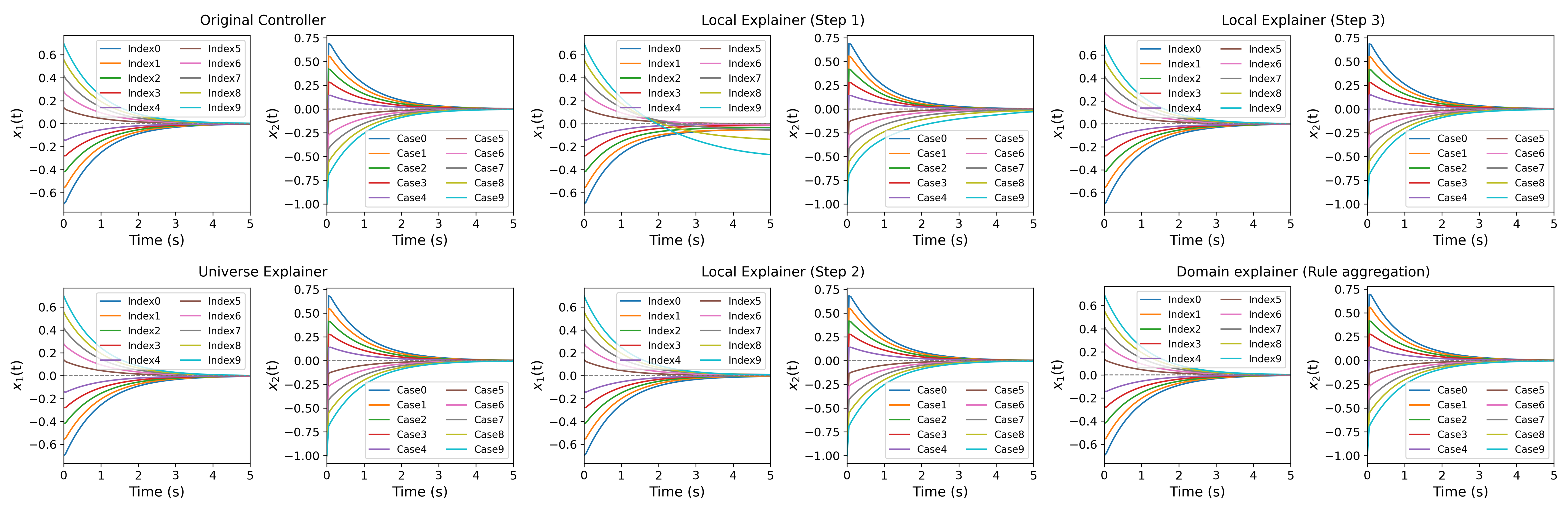}
        \centerline{(b) Domain 2}
    \end{minipage}
    
    \caption{Plots of response from the original controller and the trained explainers.} 
    \label{fig:res1}
\end{figure*}

\begin{figure*}[!htb]
    \begin{minipage}[!htb]{\linewidth}
        \centering
        \includegraphics[width=0.95\textwidth]{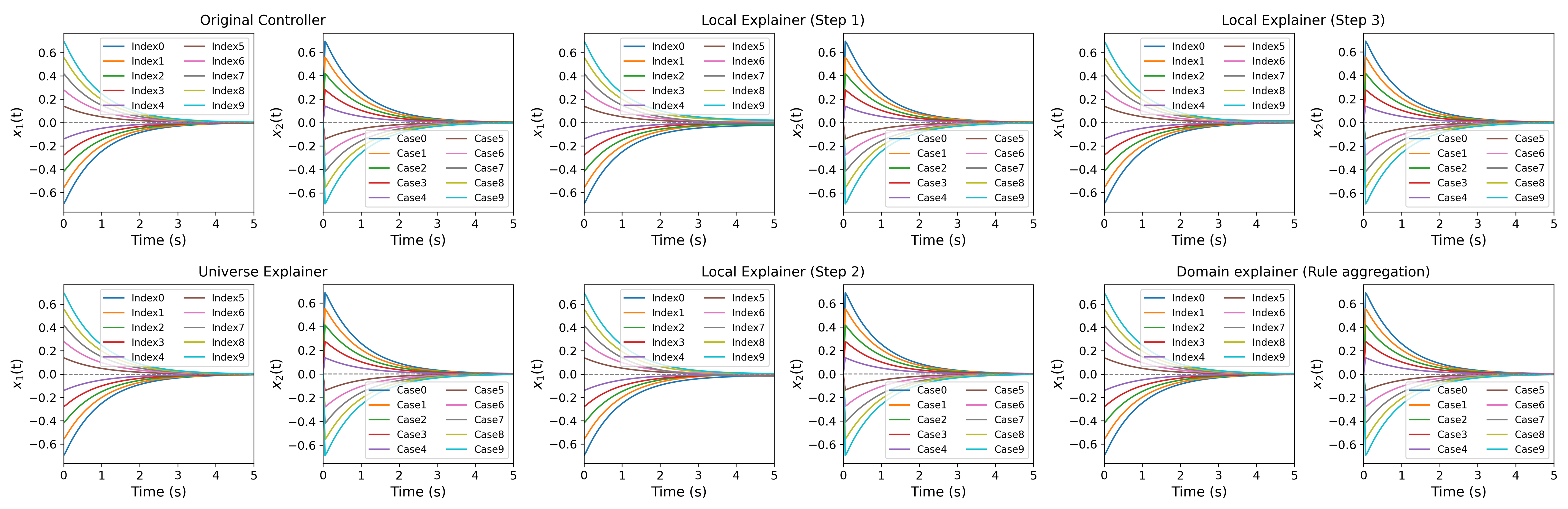}
        \centerline{(c) Domain 3}
    \end{minipage}
    \begin{minipage}[!htb]{\linewidth}
        \centering
        \includegraphics[width=0.95\textwidth]{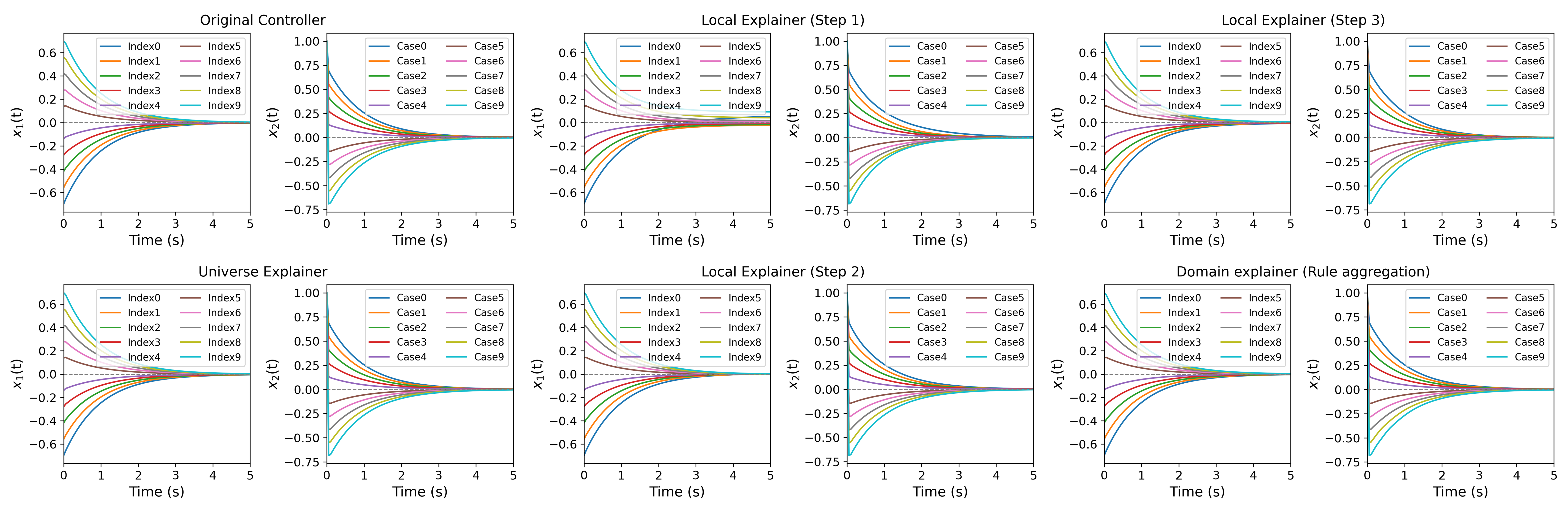}
        \centerline{(d) Domain 4}
    \end{minipage}
    \begin{minipage}[!htb]{\linewidth}
        \centering
        \includegraphics[width=0.95\textwidth]{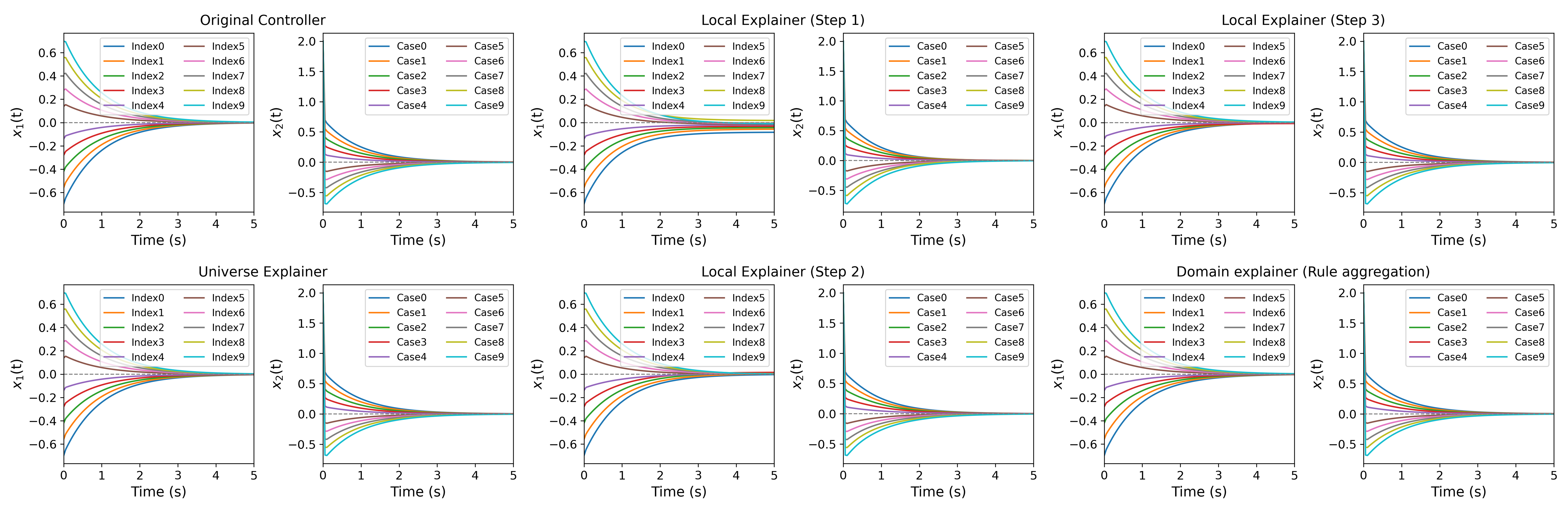}
        \centerline{(d) Domain 5}
    \end{minipage}
    \caption{Plots of response from the original controller and the trained explainers.} 
    \label{fig:res2}
\end{figure*}

Response error evaluates the deviation between the system state trajectories generated by using an explainer as the controller and those produced by the original controller, measured over the entire convergence process by solving the system dynamics. 
The response error results are reported in the last two columns of Table~\ref{tab:result1}. The response curves of the original controller and the trained explainers are presented to illustrate the dynamic behavior of different explainers in Fig.~\ref{fig:res1} and Fig.~\ref{fig:res2}.

Consistent with the observations reported in the main paper, the local explainer trained only with local input–output data (Step 1) yields larger response errors than the universe explainer. This is because it primarily captures controller behavior near the initial state, and its accuracy degrades as the system state evolves toward regions that were not sufficiently learned. This effect is also visible in Fig. \ref{fig:res1} and Fig.~\ref{fig:res2}, where the trajectory deviation of the local explainer (Step 1) increases near convergence. By incorporating global knowledge in the loss function, the local explainer (Step 2) significantly reduces the response error, producing trajectories that closely match those of the original controller. The local explainer (Step 3) further improves this alignment by explicitly minimizing response error during training, resulting in the lowest response errors among the local explainers.

For the domain explainers, weight aggregation generally preserves the response performance of the underlying local explainers, whereas the proposed rule aggregation strategy consistently achieves lower response error. As illustrated in Fig.~\ref{fig:res1} and Fig.~\ref{fig:res2}, both aggregation methods produce trajectories that closely follow the original controller across different operating regions, while rule aggregation typically provides the closest approximation.

\begin{rf}
\section{Extension of Case Study II: Turtlebot Obstacle Avoidance}

This section supplements the Turtlebot obstacle avoidance case study presented in Section V.B of the main paper. Detailed experimental settings and evaluation methods are first provided to support the quantitative comparison of explanation quality. Representative explanation results generated by XCF on two obstacle avoidance scenarios are then presented to further demonstrate the explainability of the proposed XCF on complex control systems. Finally, an illustrative scenario is presented to demonstrate how an engineer can interpret and utilize the explanations to gain insight into controller behavior

\subsection{Experimental Settings}

This subsection provides additional implementation details for the performance evaluation reported in Section V.B, including the hyperparameter settings of the explanation methods and the evaluation metrics used for quantitative comparison.

For quantitative comparison, XCF is evaluated against representative feature attribution-based XAI methods, namely SHAP, LIME and MAPLE, which have been widely adopted in explainable control and XAI studies. To ensure a fair comparison, each method was assigned a similar computational cost. The hyperparameter settings of FLS in XCF are consistent with those in inverted pendulum case in Table \ref{tab:para}, and the other settings for all methods are summarized in Table \ref{tab:turtlebot_hyper}. 

\begin{table}[h]
\centering
\caption{Hyperparameter settings for explanation methods in Case Study II.}
\resizebox{0.7\columnwidth}{!}{
\begin{tabular}{c c c}
\hline
Method & Hyperparameter & Value \\
\hline

\multirow{1}{*}{XCF}
& Number of samples & 2000 \\

\hline

\multirow{2}{*}{SHAP \cite{SHAP}}
& Max evaluations & 2000 \\
& Explainer type & Auto \\

\hline

\multirow{1}{*}{LIME \cite{LIME}}
& Number of samples & 2000 \\

\hline

\multirow{2}{*}{MAPLE \cite{MAPLE}}
& Number of estimators & 200 \\
& Number of min sample leaf & 10 \\

\hline
\end{tabular}}
\label{tab:turtlebot_hyper}
\end{table}

To quantitatively evaluate explanation quality, three metrics are adopted, namely faithfulness, monotonicity, and $R^2$ score. For explanation quality evaluation, sample points in the explanatory feature space to calculate these metrics were uniformly mapped to controller outputs using kernel regression for fair comparison across methods.

\textit{Faithfulness} \cite{faithfulness} evaluates whether the feature salience values produced by an explanation method are consistent with the actual marginal contribution of features to the controller output. It is computed as the Pearson correlation coefficient between the approximated marginal contribution vector and the feature salience vector:
\begin{equation}
{\rm Faithfulness}
=
\rho(\gamma_c,\alpha_c),
\end{equation}
where $\gamma_c$ is a $D$-dimensional vector containing the approximated marginal contribution of each feature, $\alpha_c$ denotes the feature salience vector generated by the explanation method, and $\rho(\cdot)$ represents the Pearson correlation coefficient.

\textit{Monotonicity} \cite{monotonicity} evaluates whether the feature ranking produced by an explanation method is consistent with the order of marginal improvements contributed by features. It is computed by
\begin{equation}
{\rm Monotonicity}
=
\frac{1}{D-1}
\sum^{D-2}_{d=0}
\mathbb{I}_{\delta_d > \delta_{d+1}},
\end{equation}
where $\delta_d$ denotes the marginal improvement contributed by the $d$th feature and $\mathbb{I}_{\delta_d > \delta_{d+1}}$ is an indicator function that equals 1 if the marginal improvement of the $d$th feature is greater than that of the $(d+1)$th feature, and 0 otherwise.

\textit{$R^2$ score} evaluates the fidelity of the surrogate explanation model by measuring how accurately it approximates the predictions of the original controller:
\begin{equation}
R^2
=
1-
\frac{
\sum^{M}_{m=1}
(Y_m-\hat{Y}(x_m))^2
}{
\sum^{M}_{m=1}
(Y_m-\overline{Y})^2
},
\end{equation}
where $Y_m$ and $\hat{Y}(x_m)$ denote the outputs of the original controller and the surrogate model, respectively, and $\overline{Y}$ is the average controller output.

\subsection{Extension of Simulated User Experiment}

This subsection further demonstrates the explainability of XCF on complex control systems through two representative Turtlebot obstacle avoidance scenarios. For each scenario, XCF is applied to explain the controller behavior within the key trajectory segments responsible for obstacle avoidance. 
The trajectory is first divided into multiple explanation phases according to the obstacle avoidance behavior. For each selected phase, samples are collected within the corresponding trajectory segment to train an explainer.

To focus on the turning behavior associated with obstacle avoidance, four explanatory features are considered, namely $d_{target}$, $\theta_{target}$, $d_{obs}$, and $\theta_{obs}$, representing the distance and angular deviation to the target and the nearest obstacle, respectively. For the two angular variables, a smaller value indicates that the target or obstacle is located closer to the robot's forward direction, while a larger value indicates that it is positioned further behind the robot.

For each explanation phase, an initial explainer is first trained to estimate feature salience based on the absolute salience values. The most influential two or three explanatory features are then selected to train an explainer with compact rule base for semantic inference explanation. To improve readability, two fuzzy sets are adopted when three explanatory features are selected, while three fuzzy sets are used when only two explanatory features are retained. The generated IF-THEN rules are sorted according to their decision scores in consequent in ascending order.

\begin{figure}[!htb]
    \begin{minipage}[!htb]{\linewidth}
        \centering
        \includegraphics[width=0.7\textwidth]{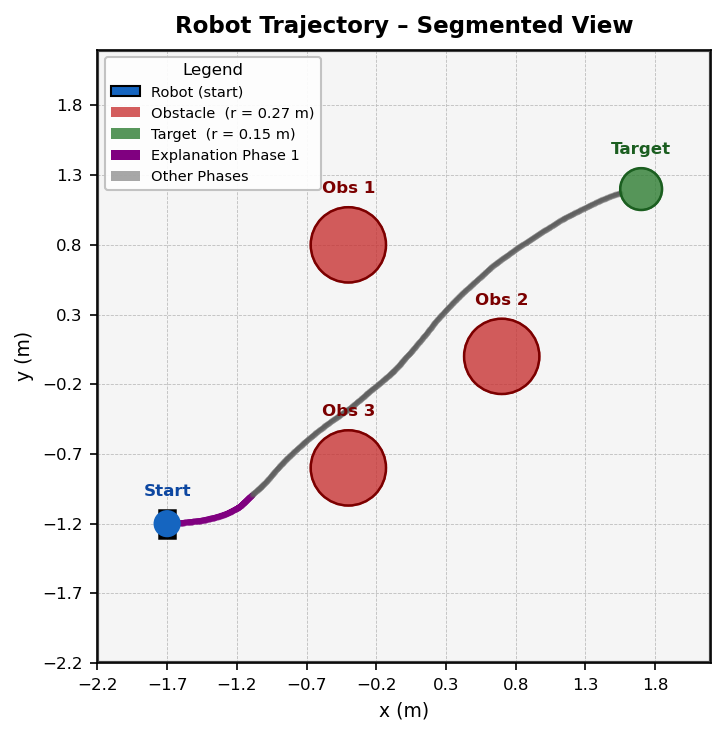}
        \centerline{(a) Case 1}
    \end{minipage}
    \begin{minipage}[!htb]{\linewidth}
        \centering
        \includegraphics[width=0.7\textwidth]{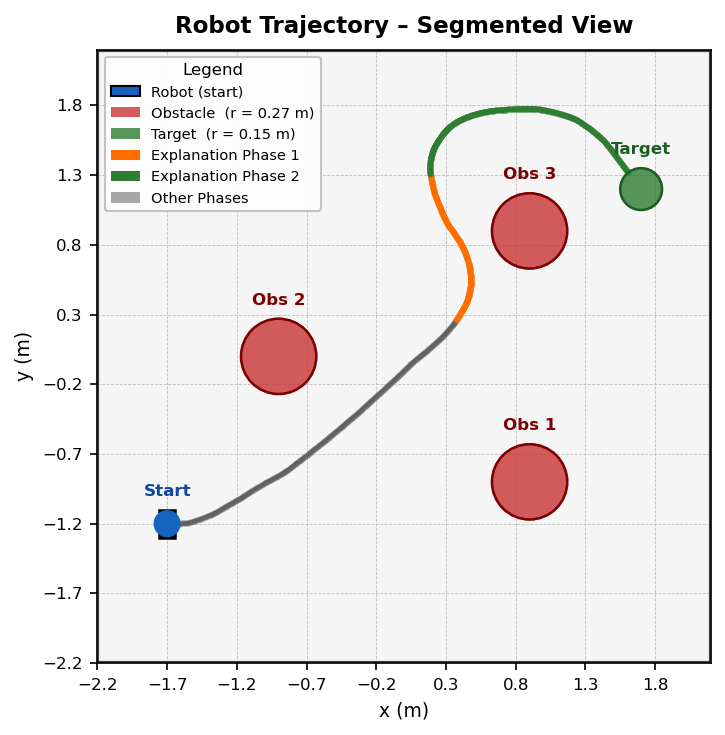}
        \centerline{(b) Case 2}
    \end{minipage}
    \caption{Trajectory segmentation of Turtlebot cases} 
    \label{fig:traj}
\end{figure}

\begin{figure}[!htb]
    \begin{minipage}[!htb]{\linewidth}
        \centering
        \includegraphics[width=0.6\textwidth]{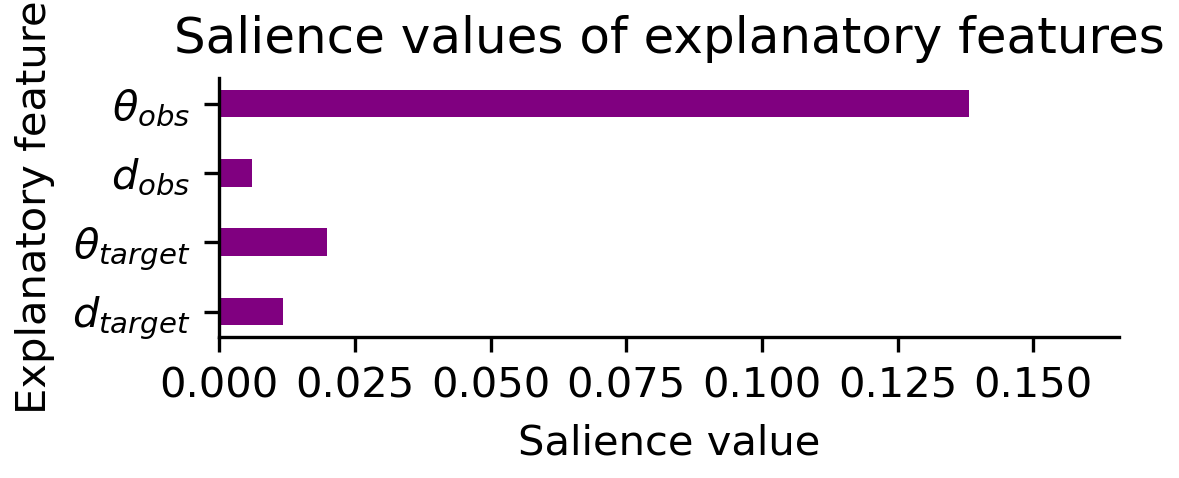}
        \centerline{(a) Salience values for Case 1}
    \end{minipage}
    \begin{minipage}[!htb]{\linewidth}
        \centering
        \includegraphics[width=\textwidth]{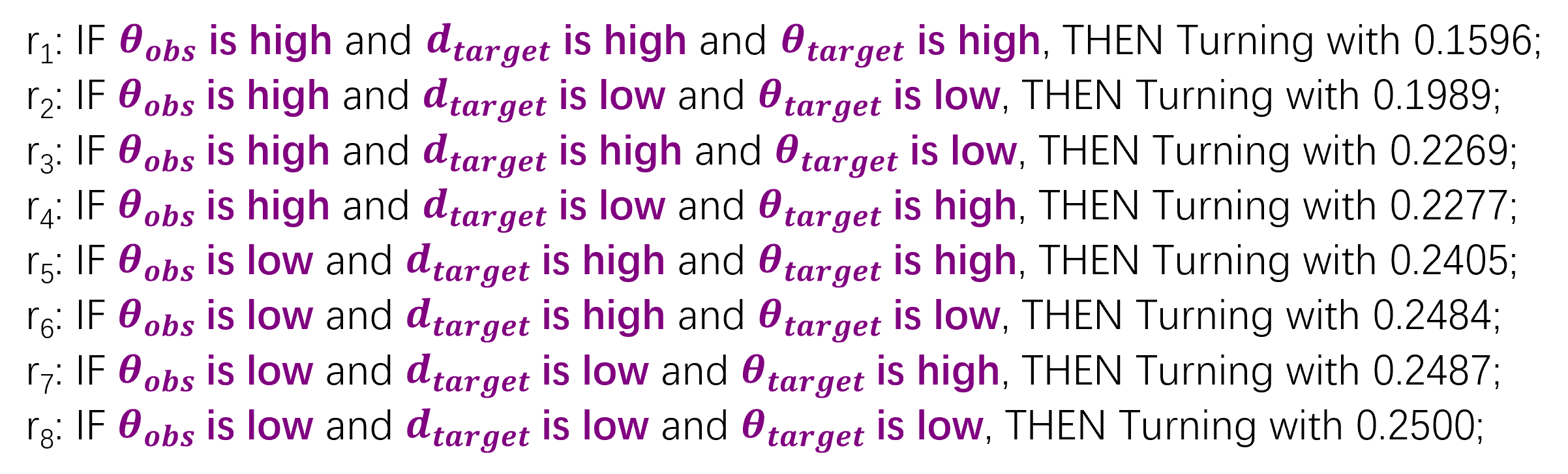}
        \centerline{(b) IF-THEN rules for Case 1}
    \end{minipage}
    \begin{minipage}[!htb]{\linewidth}
        \centering
        \includegraphics[width=0.6\textwidth]{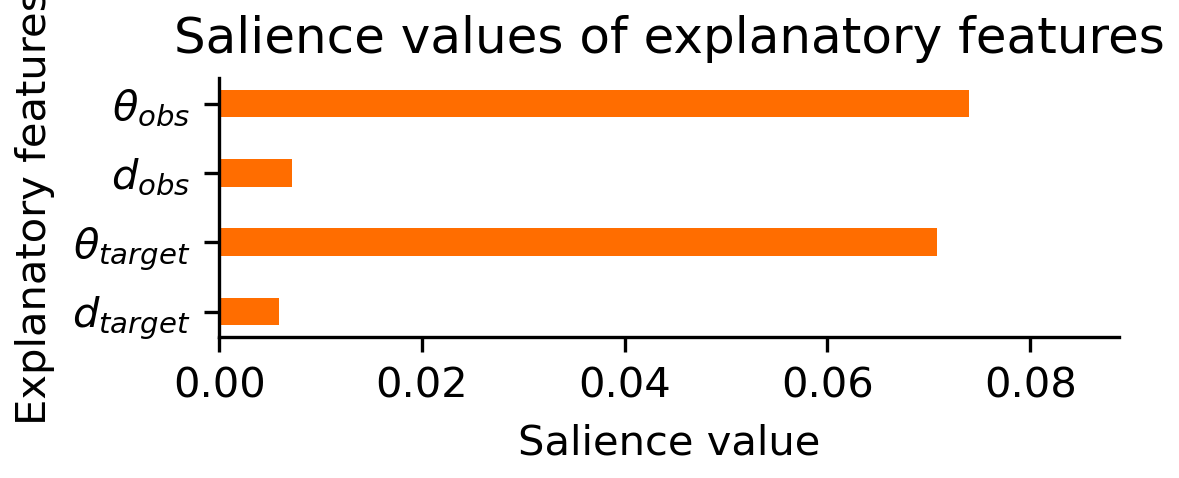}
        \centerline{(c) Salience values for Phase 1 of Case 2}
    \end{minipage}
    \begin{minipage}[!htb]{\linewidth}
        \centering
        \includegraphics[width=\textwidth]{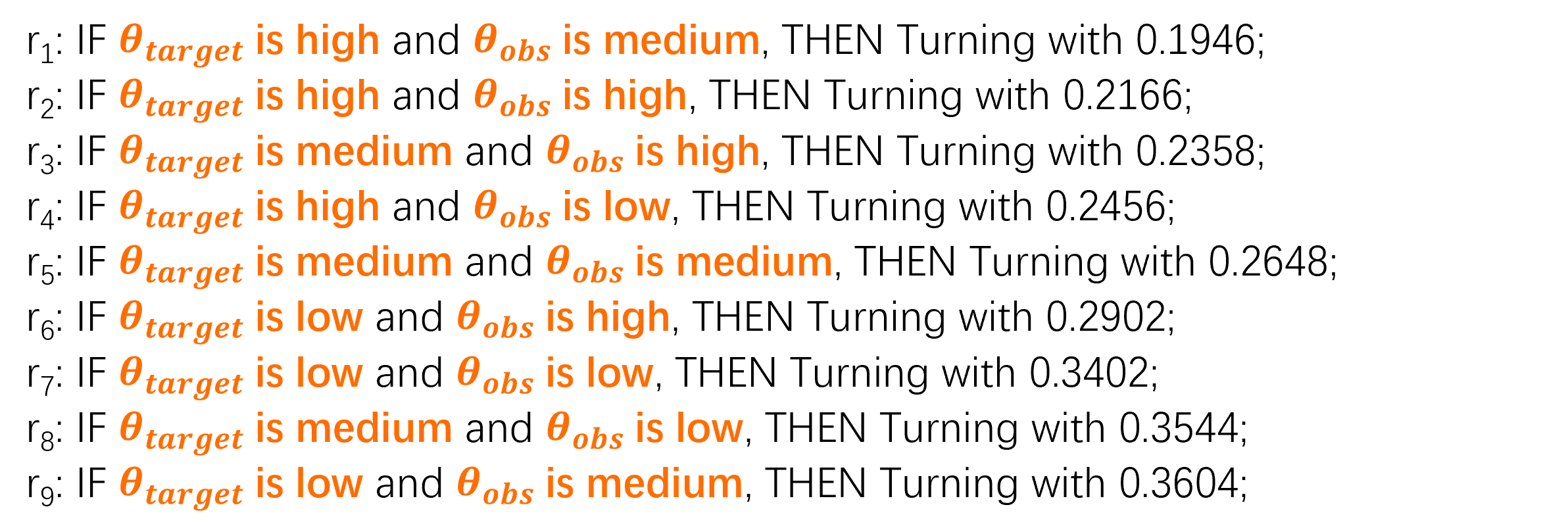}
        \centerline{(d) IF-THEN rules for Phase 1 of Case 2}
    \end{minipage}
    \begin{minipage}[!htb]{\linewidth}
        \centering
        \includegraphics[width=0.6\textwidth]{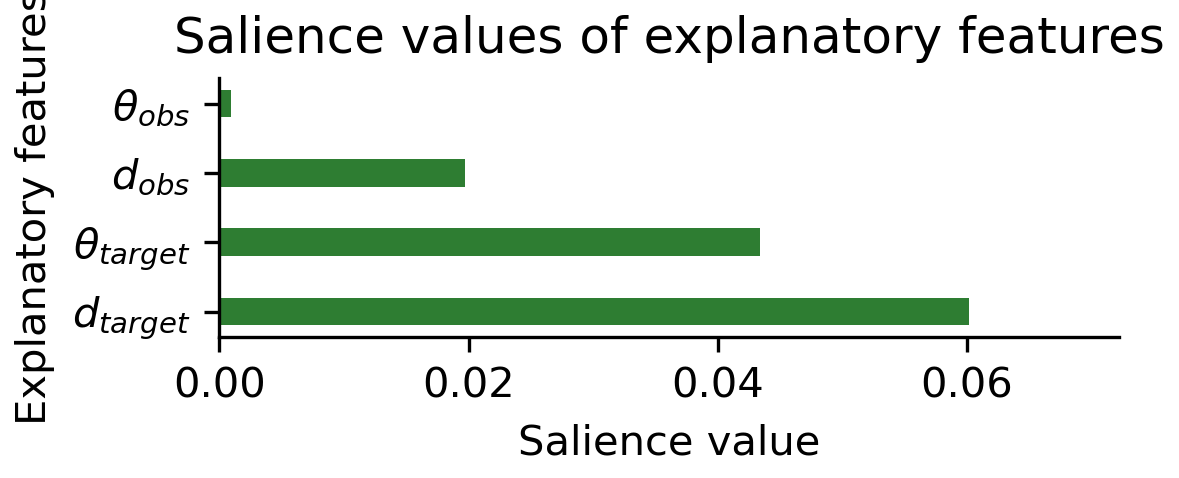}
        \centerline{(e) Salience values for Phase 2 of Case 2}
    \end{minipage}
    \begin{minipage}[!htb]{\linewidth}
        \centering
        \includegraphics[width=\textwidth]{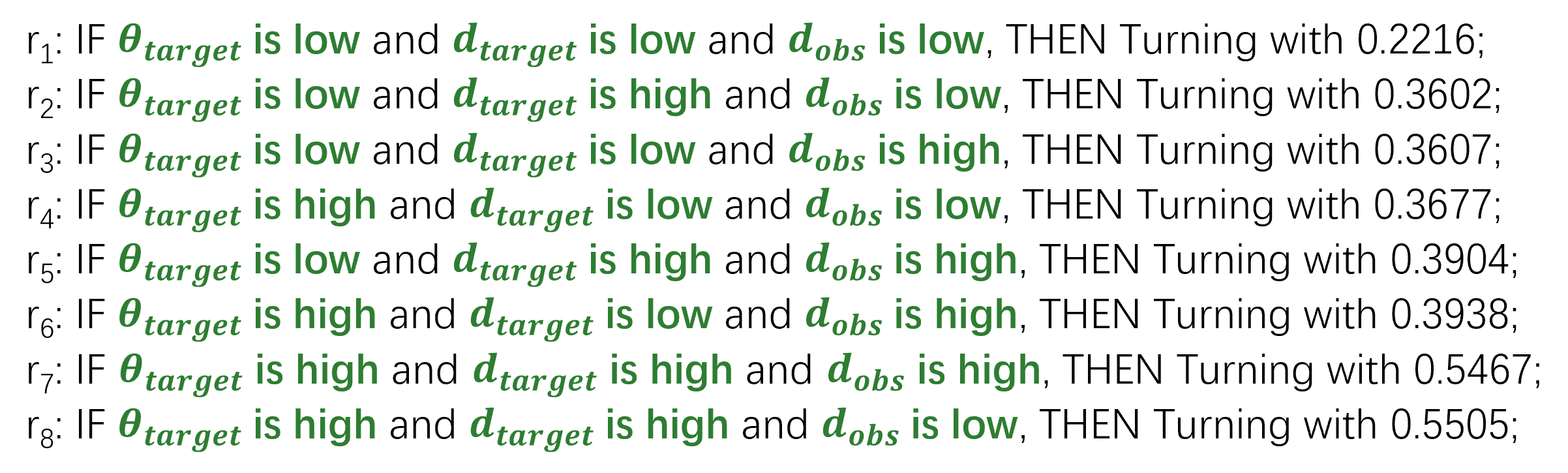}
        \centerline{(f) IF-THEN rules for Phase 2 of Case 2}
    \end{minipage}
    \caption{Explanatory interface for Turtlebot cases} 
    \label{fig:exp_s}
\end{figure}

\subsubsection{Case 1}

The trajectory of Case 1 is illustrated in Fig. \ref{fig:traj}(a). In this environment, obstacle avoidance mainly occurs during the initial stage of navigation, highlighted by the purple trajectory segment. After completing obstacle avoidance, the Turtlebot proceeds toward the target following a nearly straight path. Therefore, only the initial trajectory phase is selected for explanation.

The feature salience values in Fig. \ref{fig:exp_s}(a) reveal that three explanatory features contribute significantly to the turning behavior, among which $\theta_{obs}$ exhibits a dominant influence. In contrast, $d_{obs}$ has only a marginal contribution, suggesting that the controller mainly determines turning behavior according to the relative direction of the nearest obstacle rather than its exact distance.

The generated semantic inference rules in Fig. \ref{fig:exp_s}(b) further reveal the controller's decision logic. Since a smaller angular value indicates that the obstacle lies closer to the robot's forward direction, rules associated with lower $\theta_{obs}$ generally correspond to larger turning actions. For example, rules $r_5$--$r_8$, where $\theta_{obs}$ is low, consistently produce larger turning strengths than rules $r_1$--$r_4$, where $\theta_{obs}$ is high. This observation indicates that the controller prioritizes steering away from obstacles that are directly ahead.

Meanwhile, $d_{target}$ and $\theta_{target}$ mainly serve as secondary adjustment factors. When obstacle-related conditions remain similar, changes in target-related variables lead to only moderate variations in turning strength. This suggests that during the early obstacle avoidance stage, the controller prioritizes collision avoidance and only fine-tunes the turning behavior according to the target position.

\subsubsection{Case 2}

The environment of Case 2 differs substantially from Case 1. As illustrated in Fig. \ref{fig:traj}(b), the Turtlebot does not immediately encounter obstacles at the beginning of navigation. Instead, obstacle avoidance occurs in the later stage of the trajectory, where the target is located approximately behind an obstacle. To better explain the controller behavior, the obstacle avoidance process is divided into two explanation phases: Phase 1 corresponds to the obstacle bypassing stage (orange trajectory segment), while Phase 2 represents the target reorientation stage after clearing the obstacle (green trajectory segment).

\textit{Phase 1: Obstacle Bypassing}: 
The salience values in Fig. \ref{fig:exp_s}(c) indicate that the two angular variables, $\theta_{target}$ and $\theta_{obs}$, dominate the controller behavior, while both distance-related variables have negligible influence. This suggests that the controller mainly relies on directional information to determine turning actions during obstacle avoidance.

The generated fuzzy rules in Fig. \ref{fig:exp_s}(d) provide a more detailed explanation. Since smaller angular values indicate that the target and obstacle are located closer to the robot's forward direction, larger turning actions are observed when both $\theta_{target}$ and $\theta_{obs}$ are low or medium. For example, rules $r_7$--$r_9$ produce the largest turning strengths when both the target and obstacle are approximately aligned with the robot's heading direction. In contrast, rules $r_1$--$r_3$, where both angular variables are relatively high, correspond to smaller turning actions.

This behavior suggests that when the obstacle and target lie approximately on the same forward path, the controller intentionally increases steering intensity to deviate from the obstacle and establish a safer navigation trajectory.

\textit{Phase 2: Target Reorientation}: 
After bypassing the obstacle, the controller shifts its attention toward reaching the target. This change in behavior is clearly reflected by the salience values in Fig. \ref{fig:exp_s}(e), where $d_{target}$ and $\theta_{target}$ become the dominant explanatory features, while $\theta_{obs}$ contributes negligibly.

The generated rules in Fig. \ref{fig:exp_s}(f) indicate that stronger turning actions are produced when the Turtlebot is both far from the target and significantly misaligned with it. Specifically, rules $r_7$ and $r_8$, where both $d_{target}$ and $\theta_{target}$ are high, yield significantly larger turning strengths than the remaining rules. Conversely, when the Turtlebot is closer to the target and approximately aligned with it, such as in rule $r_1$, only a weak turning action is required.

Additionally, $d_{obs}$ still contributes to minor adjustments, indicating that obstacle information is not completely ignored even after the main obstacle avoidance stage. Overall, the explanations reveal a clear behavioral transition of the controller, from obstacle-oriented steering in Phase 1 to target-oriented navigation in Phase 2.

\subsubsection{LLM Agent-Supported User Interface}

To further enhance explanation accessibility, The explanation results of \textit{Case 2} from both Phase 1 and Phase 2, as an example, are provided to the LLM together with the controller background knowledge through a designed prompt. The complete prompt and the corresponding controller behavior explanation report generated by the LLM are provided in Tables \ref{tab:turtlebot_pmt1} and \ref{tab:turtlebot_pmt2}, and Tables \ref{tab:turtlebot_ans1}--\ref{tab:turtlebot_ans3}, respectively.

\subsection{Illustrative Scenario: Understanding and Using XCF Explanations for Turtlebot Navigation}

To demonstrate how the proposed XCF can be used in practice, this subsection presents an illustrative scenario based on the Turtlebot obstacle avoidance example. The objective is not only to show the generated explanations but also to demonstrate how an engineer can interpret and utilize them to gain insight into controller behavior.

\subsubsection{Engineering Problem}

Consider a Turtlebot operating in a navigation environment containing a target location and multiple obstacles. The robot is controlled by the FAM-HGNN controller described in Section~V-B. During testing, the robot successfully avoids collisions and reaches the target. However, the engineer observes an unexpected behavior: the robot suddenly performs a sharp turning maneuver and temporarily moves away from the direct path to the target.

Although the trajectory can be visualized and the wheel commands can be recorded, these observations alone cannot explain why the controller made this decision. Since the controller is represented by a deep reinforcement learning architecture incorporating graph neural networks and fuzzy attention mechanisms, analyzing the internal network parameters provides little intuitive understanding.

The engineer therefore seeks answers to the following questions:

\begin{itemize}
\item Why did the robot perform the turning maneuver?
\item Was the maneuver caused by obstacle avoidance or target tracking?
\item Which environmental factors contributed most to the decision?
\item Is the behavior reasonable or does it indicate a potential controller issue?
\item How can the obtained explanation be used to improve future controller designs?
\end{itemize}

The proposed XCF addresses these questions by transforming the original black-box controller behavior into interpretable explanations consisting of salience values, semantic rules, and natural-language descriptions.

\subsubsection{Step 1: Understanding What the Controller Cares About}

Rather than directly explaining the original high-dimensional robot state, XCF employs semantically meaningful explanatory variables:

\begin{itemize}
\item Distance to target ($d_{target}$)
\item Target angle ($\theta_{target}$)
\item Distance to nearest obstacle ($d_{obs}$)
\item Obstacle angle ($\theta_{obs}$)
\end{itemize}

These quantities correspond closely to how a human operator would naturally describe the navigation situation.

For the obstacle bypassing phase (Phase~1), the computed salience values (by Eq. (10) in the main text) in Fig. \ref{fig:exp_s}(c) are
\begin{align}
\theta_{obs} = 0.0740, \nonumber\
\theta_{target} = 0.0708, \nonumber\
\\
d_{obs} = 0.0071, \nonumber\
d_{target} = 0.0059.
\end{align}
It can be observed that 
$[
\theta_{obs}\approx\theta_{target}
\gg
d_{obs},d_{target}
].$ 
This indicates that the controller primarily relies on directional information rather than distance information.

From an engineering perspective, this means that the robot is paying much more attention to where the obstacle and target are located relative to its heading than to how far away they are.

Without XCF, the engineer only observes:

\begin{quote}
``The robot turned.''
\end{quote}

With XCF, the engineer learns:

\begin{quote}
``The robot turned because the relative directions of the obstacle and target became important.''
\end{quote}

\subsubsection{Step 2: Understanding How the Controller Makes Decisions}

The salience values indicate which variables are important, but they do not explain how those variables are used. 
To reveal the decision logic, XCF generates semantic IF--THEN rules (THEN part are computed by Eq. (11) in the main text).

For Phase~1, representative rules in Fig.\ref{fig:exp_s}(d) include:
\begin{align}
r_1:&\ \text{High }\theta_{target}, \nonumber\
&\ \text{Medium }\theta_{obs} \nonumber\
&\ \rightarrow 0.1946,
\\
r_7:&\ \text{Low }\theta_{target}, \nonumber\
&\ \text{Low }\theta_{obs}, \nonumber\
&\ \rightarrow 0.3402,
\\
r_8:&\ \text{Medium }\theta_{target}, \nonumber\
&\ \text{Low }\theta_{obs}, \nonumber\
&\ \rightarrow 0.3544,
\\
r_9:&\ \text{Low }\theta_{target}, \nonumber\
&\ \text{Medium }\theta_{obs}, \nonumber\
&\ \rightarrow 0.3604.
\end{align}

The generated rules reveal that lower values of $\theta_{obs}$ consistently lead to larger turning strengths.

In simple terms, when the obstacle lies closer to the robot's forward direction, the controller produces a stronger steering command. 
The engineer therefore gains the following insight:

\begin{quote}
``The sharp turn is not random. The controller intentionally increases steering effort when an obstacle blocks the forward path.''
\end{quote}

This explanation provides significantly more understanding than simply observing the resulting trajectory.

\subsubsection{Step 3: Understanding Behavioral Strategy Changes}

The trajectory is divided into two explanation phases.

\paragraph{Phase 1: Obstacle Bypassing}

During obstacle bypassing, obstacle-related information dominates the decision process. 
The explanation therefore reveals that the controller is operating in a safety-oriented mode:

\begin{quote}
``Avoid collision first.''
\end{quote}

\paragraph{Phase 2: Target Reorientation}

After the obstacle has been cleared, the salience values become
\begin{align}
d_{target} &= 0.0602, \nonumber\
\theta_{target} &= 0.0434, \nonumber\
\\
d_{obs} &= 0.0198, \nonumber\
\theta_{obs} &= 0.0009.
\end{align}
Notice that 
$[
\theta_{obs}=0.0009
]$, 
which is almost negligible. 
This indicates that obstacle direction no longer influences the controller significantly.

Instead, target-related information dominates the decision process. 
Representative semantic rules include
\begin{align}
r_1:&\ \text{Low }\theta_{target}, \nonumber\
&\ \text{Low }d_{target}, \nonumber\
&\ \text{Low }d_{obs} \nonumber\
&\ \rightarrow 0.2216,
\\
r_7:&\ \text{High }\theta_{target}, \nonumber\
&\ \text{High }d_{target}, \nonumber\
&\ \text{High }d_{obs} \nonumber\
&\ \rightarrow 0.5467,
\\
r_8:&\ \text{High }\theta_{target}, \nonumber\
&\ \text{High }d_{target}, \nonumber\
&\ \text{Low }d_{obs} \nonumber\
&\ \rightarrow 0.5505.
\end{align}
These rules indicate that stronger turning actions are generated when the robot is poorly aligned with the target and still far away from the desired destination.

The engineer therefore obtains the following insight:

\begin{quote}
``The controller has shifted from obstacle avoidance to target reorientation.''
\end{quote}

\subsubsection{Step 4: Reconstructing the Controller Reasoning Process}

By combining salience values and semantic rules, the engineer can reconstruct the controller's decision-making process:

\begin{enumerate}
\item The obstacle enters the robot's forward direction.
\item Obstacle angle becomes a dominant explanatory factor.
\item Strong avoidance rules are activated.
\item A large turning maneuver is generated.
\item The robot safely bypasses the obstacle.
\item Obstacle influence decreases.
\item Target-related variables become dominant.
\item The robot gradually returns toward the target.
\end{enumerate}

This explanation reveals not only what happened but also why it happened.

\subsubsection{Role of the LLM-Based Explanation Interface}

While salience values and semantic rules provide interpretable information, engineers may still need to inspect multiple figures and rule tables to understand the overall behavior.

The LLM-based explanation interface further transforms the generated explanations into concise natural-language descriptions. 
For example, instead of presenting: 

$[
\theta_{obs}=0.0740,
\qquad
\theta_{target}=0.0708,
]$ 

together with multiple semantic rules, the LLM can generate a summary such as:

\begin{quote}
``The robot is primarily reacting to obstacle direction. Because the obstacle lies close to the robot's forward path, strong avoidance rules are activated, resulting in a sharp turning maneuver. After clearing the obstacle, the controller shifts its attention toward the target and resumes goal-oriented navigation.''
\end{quote}

This enables users without expertise in fuzzy systems, reinforcement learning, or control theory to understand the controller behavior.

\subsubsection{Engineering Impact and Transformation}

The generated explanations provide benefits beyond simple interpretability.

\begin{itemize}
    \item \textit{Controller Validation}: Engineers can verify whether the learned policy follows the intended navigation strategy.
    \item \textit{Fault Diagnosis}: Unexpected salience distributions or abnormal rules can help identify controller design issues, reward-design problems, or undesirable learning outcomes.
    \item \textit{Controller Improvement}: The explanations reveal which factors dominate decision making. This information can guide modifications to reward functions, attention mechanisms, safety constraints, and controller architectures.
    \item \textit{Trust and Safety Assessment}: The explanations demonstrate that the controller is not behaving randomly but is following an interpretable sequence of decision-making steps.
\end{itemize}

Consequently, XCF transforms controller evaluation from merely observing robot behavior to understanding the reasoning underlying that behavior. This deeper understanding supports debugging, validation, safety assessment, certification, and the development of more trustworthy autonomous systems.
\end{rf}

\begin{table*}[ht]
\centering
\caption{Prompt and Answer of Inverted Pendulum Case on Local Level}
\begin{tabular}{p{0.95\textwidth}}
\hline
\textbf{Interpreter Prompt (Markdown format)}:  \\ 
\hline
Please generate a brief Controller Behavior Explanation Report based on the provided background knowledge of the controller and the structured explanation results from upstream XAI algorithms. The XAI methods have generated the explanations for the controller behavior on local level for a specific initial state. There are two types of explanations for the controller's behavior: salience values reflecting the contribution of each system state variable to the controller's output, and IF-THEN rules reflecting the decision logic of the controller. \\ 
 \\ 
\#\# Background Knowledge of the Controller: \\ 
The controller is for stabilizing a cart-pole typed inverted pendulum system, where x1 denotes the angular displacement of the pendulum,
x2 denotes the angular velocity of the pendulum, and u denotes the force applied to the cart
(For x1 and x2, positive value means anti-clockwise direction; for u, positive value means leftward force).
 \\ 
\#\# Explanation results from XAI algorithms: \\ 
\#\#\# Salience Value of State Variables Table \\ 
$|$ System State $|$ Salience Value $|$ \\ 
$|$------------------$|$--------------------$|$ \\ 
$|$ x2 $|$ 0.436 $|$ \\ 
$|$ x1 $|$ 0.070 $|$ \\ 

\#\#\# IF-THEN Rules of Decision Logic: \\ 
r1: IF x2 is relatively low, x1 is relatively low, THEN u is 3437.476; \\
r2: IF x2 is relatively low, x1 is relatively neutral, THEN u is 3760.088; \\
r3: IF x2 is relatively low, x1 is relatively high, THEN u is 4095.529; \\
r4: IF x2 is relatively neutral, x1 is relatively low, THEN u is 5286.157; \\
r5: IF x2 is relatively neutral, x1 is relatively neutral, THEN u is 5618.134; \\
r6: IF x2 is relatively neutral, x1 is relatively high, THEN u is 5927.711; \\
r7: IF x2 is relatively high, x1 is relatively low, THEN u is 7204.669; \\
r8: IF x2 is relatively high, x1 is relatively neutral, THEN u is 7482.967; \\
r9: IF x2 is relatively high, x1 is relatively high, THEN u is 7744.112; \\
 \\ 
\#\# Task: \\ 
Generate a concise, four-paragraph Controller Behavior Explanation Report using the provided explanation results with the following structure: \\ 

$'''$ \\ 
\# Controller Behavior Explanation Report \\ 
1. Paragraph 1 (Introduction): Briefly summarize the purpose of the report and the type of explanation results being analyzed. Mention that the explanations aim to reveal the internal logic of a black-box controller based on its input-output behavior. \\ 
2. Paragraph 2 (Salience Analysis): Analyze the salience values. Identify which system state variables have the highest and lowest influence on the controller's output. Discuss the relative importance of these variables and interpret what this suggests about the controller's decision-making focus. \\ 
3. Paragraph 3 (Rule Analysis): Analyze the IF-THEN rules. Group rules by the similarity of their THEN-part outputs. Summarize the typical conditions in the IF-part that lead to each type of control action. Highlight any consistent patterns in how combinations of state variable values lead to specific outputs. \\ 
4. Paragraph 4 (Conclusion): Provide a brief summary of the controller’s behavior based on the salience and rule analyses. Reflect on what the explanation reveals about the controller's strategy in handling different system states. \\ 
$'''$ \\ 
 \\ 
\#\# Requirements: \\ 
- The report should be written in a clear and professional tone, suitable for an academic or engineering audience. \\ 
- Do not include section or paragraph headings; write as continuous prose with logical transitions between paragraphs.
\\ \hline
\textbf{Interpreter Answer}:  \\ 
\hline
\textbf{{\normalsize Controller Behavior Explanation Report}} \\ 
The purpose of this report is to provide a concise explanation of the controller’s behavior for a cart-pole inverted pendulum system, based on local explanation results produced by upstream explainable AI (XAI) methods. These explanations are derived from the controller’s observed input–output behavior at a specific initial state and aim to uncover the internal decision logic of an otherwise black-box controller. Two complementary explanation forms are analyzed: salience values that quantify the relative influence of each state variable on the control output, and a set of IF–THEN rules that describe how combinations of system states map to specific control actions.

The salience analysis indicates that the angular velocity of the pendulum, denoted by $x_2$, has the highest influence on the control force, with a salience value of $0.436$, while the angular displacement, $x_1$, has a much lower influence, with a salience value of $0.070$. This large disparity suggests that the controller prioritizes dynamic information about how fast the pendulum is moving over static information about its current angle. In practical terms, the controller appears to focus primarily on damping motion and counteracting rapid changes, using angular position as a secondary refinement rather than the main driver of its decisions.

The IF--THEN rules can be grouped by the magnitude of the resulting control force $u$. When $x_2$ is relatively low, the controller applies comparatively small forces (approximately $3{,}400$--$4{,}100$), with the force increasing as $x_1$ transitions from low to high. For relatively neutral values of $x_2$, the control force shifts to a medium range (about $5{,}300$--$5{,}900$), again increasing monotonically with higher $x_1$. When $x_2$ is relatively high, the controller consistently produces the largest forces (around $7{,}200$--$7{,}700$), with further increments corresponding to higher $x_1$. Across all groups, a consistent pattern emerges: increases in either angular velocity or angular displacement lead to stronger corrective actions, with changes in $x_2$ producing the most pronounced jumps in output.

Overall, the combined salience and rule analyses reveal a controller strategy that is strongly velocity-driven and structured in a systematic, monotonic manner. The controller responds aggressively to high angular velocities to quickly suppress unstable motion, while angular displacement modulates the force within each velocity regime. This behavior is consistent with a stabilization strategy that emphasizes rapid damping of pendulum motion while still accounting for positional deviation, thereby maintaining balance across a wide range of system states.

\\ \hline
\end{tabular}
\label{tab:local}
\end{table*}

\begin{table*}[ht]
\centering
\caption{Prompt of Inverted Pendulum Case on Domain Level}
\begin{tabular}{p{0.95\textwidth}}
\hline
\textbf{Interpreter Prompt (Markdown format)}:  \\ 
\hline
Please generate a comprehensive Controller Behavior Explanation Report based on the provided background knowledge of the controller and the structured explanation results from upstream XAI algorithms. The XAI methods have generated the explanations for the controller behavior on domain level according to the user-specified region in the state space. There are two types of explanations for the controller's behavior: salience values reflecting the contribution of each system state variable to the controller's output, and IF-THEN rules reflecting the decision logic of the controller. \\
 \\
\#\# Background Knowledge of the Controller: \\
The controller is for stabilizing a cart-pole typed inverted pendulum system, where x1 denotes the angular displacement of the pendulum,
x2 denotes the angular velocity of the pendulum, and u denotes the force applied to the cart
(For x1 and x2, positive value means anti-clockwise direction; for u, positive value means leftward force). \\
 \\
\#\# Explanation results from XAI algorithms: \\
\#\#\# Salience Value of State Variables Table \\
$|$ System State $|$ Salience Value $|$ \\
$|$------------------$|$--------------------$|$ \\
$|$ x2 $|$ 0.437 $|$ \\
$|$ x1 $|$ 0.070 $|$ \\

\#\#\# IF-THEN Rules of Decision Logic: \\
r1: IF x2 is relatively low, x1 is relatively low, THEN u is 4150.866; \\
r2: IF x2 is relatively low, x1 is relatively neutral, THEN u is 4380.113; \\
r3: IF x2 is relatively low, x1 is relatively high, THEN u is 4610.267; \\
r4: IF x2 is relatively neutral, x1 is relatively low, THEN u is 5987.736; \\
r5: IF x2 is relatively neutral, x1 is relatively neutral, THEN u is 6216.830; \\
r6: IF x2 is relatively neutral, x1 is relatively high, THEN u is 6446.559; \\
r7: IF x2 is relatively high, x1 is relatively low, THEN u is 7820.796; \\
r8: IF x2 is relatively high, x1 is relatively neutral, THEN u is 8050.345; \\
r9: IF x2 is relatively high, x1 is relatively high, THEN u is 8281.248; \\
 \\
\#\# Task: \\
Generate a professional Controller Behavior Explanation Report with the following structure: \\

$'''$ \\
\# Controller Behavior Explanation Report \\

\#\# Summary of Controller Behavior \\
Provide a brief introduction of the following subsections. \\

\#\#\# State Contribution Analysis \\
Analyze the system state variable significance values table to explain: \\
- Which state variables have the most significant impact on the controller's output \\
- The relative importance of each variable \\
- How the significance values indicate the contribution of each variable to the control decisions \\

\#\#\# Controller Logic Analysis \\
Analyze the IF-THEN rules table to explain: \\
- The key premises or conditions that govern the controller's decision-making process \\
(Hint: Try to group the rules with similar outputs and identify the common premises in IF parts for each group) \\
- How these rules work together to form the overall control strategy \\
- Any notable rule interactions or hierarchies \\

\#\# Conclusion \\
Provide a concise summary that: \\
- Synthesizes the insights from both the variable influence analysis and logic rule analysis \\
- Highlights the most critical factors that drive the controller's decisions \\
- Explains the overall behavior characteristics of the controller \\
$'''$ \\
 \\
\#\# Requirements: \\
- Use clear, professional English suitable for technical documentation \\
- Provide specific references to the salience values and rules when making explanations \\
- Ensure the report is comprehensive yet accessible to users who need to understand the controller's behavior \\
- Focus on practical insights that help users understand why and how the controller makes decisions
\\ \hline
\end{tabular}
\label{tab:domain_pmt}
\end{table*}

\begin{table*}[ht]
\centering
\caption{Answer of Inverted Pendulum Case on Domain Level}
\begin{tabular}{p{0.95\textwidth}}
\hline
\textbf{Interpreter Answer}: \\ 
\hline
\textbf{\normalsize Controller Behavior Explanation Report} \\

\textbf{Summary of Controller Behavior} \\
This report explains the behavior of a controller designed to stabilize a cart--pole type inverted pendulum system. The explanation integrates two complementary perspectives provided by upstream XAI algorithms: (i) a \emph{state contribution analysis} based on salience values, which quantifies the relative influence of each system state on the controller’s output, and (ii) a \emph{controller logic analysis} based on extracted IF--THEN rules, which reveals the qualitative decision-making structure used to determine the control force~$u$. Together, these analyses provide both quantitative and qualitative insight into how and why the controller generates its control actions. \\
 \\
\textbf{State Contribution Analysis} \\
The salience values indicate how strongly each state variable contributes to the controller’s output~$u$:

\begin{itemize}
    \item The angular velocity of the pendulum~($x_2$) has a salience value of~$0.437$, making it the most influential state variable.
    \item The angular displacement of the pendulum~($x_1$) has a salience value of~$0.070$, indicating a much smaller, though non-negligible, contribution.
\end{itemize}

The large disparity between these values suggests that the controller primarily reacts to the \emph{rate of change of the pendulum angle} rather than the angle itself. In practical terms, this means the controller is highly sensitive to how fast the pendulum is rotating (clockwise or anti-clockwise) and uses this information as the dominant signal to compute the cart force. The pendulum’s angular displacement plays a secondary, fine-tuning role, adjusting the control force once the overall response dictated by angular velocity is determined. \\

This prioritization is consistent with stabilization objectives in inverted pendulum systems: rapid changes in angular velocity can quickly destabilize the system, so the controller emphasizes damping and corrective action based on~$x_2$, while~$x_1$ refines the response to ensure accurate positioning around the upright equilibrium. \\
 \\
\textbf{Controller Logic Analysis} \\
The IF--THEN rules (r1--r9) describe how combinations of angular velocity~$x_2$ and angular displacement~$x_1$ map to specific control forces~$u$. These rules can be grouped and interpreted as follows. \\

1) Dominant role of angular velocity \\
The rules naturally cluster into three groups based on the qualitative level of~$x_2$:

\begin{itemize}
    \item \textbf{Low $x_2$} (r1--r3): control forces range from approximately $4150.866$ to $4610.267$.
    \item \textbf{Neutral $x_2$} (r4--r6): control forces increase to a mid-range, from $5987.736$ to $6446.559$.
    \item \textbf{High $x_2$} (r7--r9): control forces are the largest, from $7820.796$ to $8281.248$.
\end{itemize}

This progression shows a clear hierarchy: as the angular velocity increases from low to high, the controller systematically applies a stronger leftward force~$u$. This reflects a strategy of increasingly aggressive intervention to counteract faster pendulum motion. \\

2) Modulating role of angular displacement \\
Within each~$x_2$ group, the three corresponding rules differ only in the level of~$x_1$ (relatively low, neutral, or high). For a fixed~$x_2$ level, the control force~$u$ increases monotonically as~$x_1$ increases (e.g., r1 $\rightarrow$ r2 $\rightarrow$ r3, or r7 $\rightarrow$ r8 $\rightarrow$ r9). This indicates that angular displacement acts as a \emph{secondary adjustment mechanism}, refining the control effort based on how far the pendulum has deviated from the upright position. \\

3) Overall control strategy and rule interaction \\
Taken together, the rules implement a structured and hierarchical strategy. First, the controller assesses the \emph{severity of motion} via~$x_2$, which determines the baseline magnitude of the control force. Second, this baseline is adjusted according to the current angular deviation~$x_1$, ensuring precise stabilization. \\

There are no conflicting rules; instead, they form a smooth, ordered decision surface over the~$(x_1, x_2)$ state space. This consistency suggests that the controller behavior is well coordinated, with angular velocity driving the main response and angular displacement providing incremental corrections. \\
 \\
\textbf{Conclusion} \\
By combining the salience value analysis and the IF--THEN rule analysis, a coherent picture of the controller’s behavior emerges. The controller’s decisions are driven primarily by the pendulum’s angular velocity~($x_2$), as evidenced by its high salience value~($0.437$) and by the rule structure that scales control force significantly with changes in~$x_2$. Angular displacement~($x_1$), with a much lower salience value~($0.070$), serves as a secondary factor that fine-tunes the applied force within each angular velocity regime. \\

Overall, the controller exhibits a \emph{velocity-dominated, hierarchically structured control strategy}: it reacts strongly to fast pendulum motion to ensure stability, while using angular position information to refine and smooth the control action. This behavior is well aligned with the practical requirements of stabilizing an inverted pendulum, providing both robustness against rapid disturbances and precision near the equilibrium point.

\\ \hline
\end{tabular}
\label{tab:domain_ans}
\end{table*}

\begin{table*}[ht]
\centering
\caption{Prompt of Turtlebot Obstacle Avoidance Case (Part 1)}
\begin{tabular}{p{0.95\textwidth}}
\hline
\textbf{Interpreter Prompt (Markdown format with Fig. \ref{fig:traj})}:  \\ 
\hline
Please generate a comprehensive Controller Behavior Explanation Report based on the provided background knowledge of the controller and the structured explanation results from upstream XAI algorithms. The XAI methods have generated explanations for the controller behavior at the domain level according to user-specified trajectory segments during obstacle avoidance. There are two types of explanations for the controller's behavior: salience values reflecting the contribution of explanatory variables to the controller's turning behavior, and IF-THEN rules reflecting the controller's decision logic. \\
\\
\#\# Background Knowledge of the Controller:

The controller is designed for a Turtlebot obstacle avoidance navigation task. The Turtlebot navigates from a start position to a target position while avoiding multiple static obstacles.

To improve interpretability, the original high-dimensional observation space is mapped into an explanatory feature space. The controller behavior is explained using the following explanatory variables:

- $d_{{target}}$: distance between the Turtlebot and the target position

- $\theta_{{obs}}$: angular deviation between the Turtlebot's heading direction and the target position

- $d_{{obs}}$: distance between the Turtlebot and the nearest obstacle

- $\theta_{{obs}}$: angular deviation between the Turtlebot's heading direction and the nearest obstacle

For the angular variables:
- Smaller values indicate that the target or obstacle is closer to the Turtlebot's forward direction
- Larger values indicate that the target or obstacle is positioned further behind the Turtlebot

The explained controller output is the turning behavior of the Turtlebot. Larger output values indicate stronger turning actions. \\
\\

\#\# Background of the Current Explanation Phases:

In this case, obstacle avoidance occurs in the later stage of the trajectory. The target is located directly behind the obstacle encountered in the current forward direction, and it is on the same straight line as the obstacle.

To better explain the controller behavior, the obstacle avoidance process is divided into two explanation phases:

- Phase 1: Obstacle bypassing stage. During this phase, the Turtlebot actively steers away from the obstacle to establish a safe navigation path (see the uploaded figure of trajectory segmentation).

- Phase 2: Target reorientation stage. After bypassing the obstacle, the Turtlebot adjusts its heading direction back toward the target position.

The explanation results therefore characterize the controller's decision logic during these two consecutive navigation phases. \\
\\

\#\# Explanation Results from XAI Algorithms:

\#\#\# Phase 1: Obstacle Bypassing Stage

\#\#\#\# Salience Value of State Variables Table

$|$ System State $|$ Salience Value $|$

$|$---------------$|$----------------$|$

$|$ $\theta_{{obs}}$ $|$ 0.0740 $|$

$|$ $\theta_{{target}}$ $|$ 0.0708 $|$

$|$ $d_{{obs}}$ $|$ 0.0071 $|$

$|$ $d_{{target}}$ $|$ 0.0059 $|$

\#\#\#\# IF-THEN Rules of Decision Logic:

r1: IF $\theta_{{target}}$ is high, $\theta_{{obs}}$ is medium, THEN Turning is 0.1946;

r2: IF $\theta_{{target}}$ is high, $\theta_{{obs}}$ is high, THEN Turning is 0.2166;

r3: IF $\theta_{{target}}$ is medium, $\theta_{{obs}}$ is high, THEN Turning is 0.2358;

r4: IF $\theta_{{target}}$ is high, $\theta_{{obs}}$ is low, THEN Turning is 0.2456;

r5: IF $\theta_{{target}}$ is medium, $\theta_{{obs}}$ is medium, THEN Turning is 0.2648;

r6: IF $\theta_{{target}}$ is low, $\theta_{{obs}}$ is high, THEN Turning is 0.2902;

r7: IF $\theta_{{target}}$ is low, $\theta_{{obs}}$ is low, THEN Turning is 0.3402;

r8: IF $\theta_{{target}}$ is medium, $\theta_{{obs}}$ is low, THEN Turning is 0.3544;

r9: IF $\theta_{{target}}$ is low, $\theta_{{obs}}$ is medium, THEN Turning is 0.3604; \\
\\
\#\#\# Phase 2: Target Reorientation Stage

\#\#\#\# Salience Value of State Variables Table

$|$ System State $|$ Salience Value $|$

$|$---------------$|$----------------$|$

$|$ $d_{{target}}$ $|$ 0.0602 $|$

$|$ $\theta_{{target}}$ $|$ 0.0434 $|$

$|$ $d_{{obs}}$ $|$ 0.0198 $|$

$|$ $\theta_{{obs}}$ $|$ 0.0009 $|$

\#\#\#\# IF-THEN Rules of Decision Logic:

r1: IF $\theta_{{target}}$ is low, $d_{{target}}$ is low, $d_{{obs}}$ is low, THEN Turning is 0.2216;

r2: IF $\theta_{{target}}$ is low, $d_{{target}}$ is high, $d_{{obs}}$ is low, THEN Turning is 0.3602;

r3: IF $\theta_{{target}}$ is low, $d_{{target}}$ is low, $d_{{obs}}$ is high, THEN Turning is 0.3607;

r4: IF $\theta_{{target}}$ is high, $d_{{target}}$ is low, $d_{{obs}}$ is low, THEN Turning is 0.3677;

r5: IF $\theta_{{target}}$ is low, $d_{{target}}$ is high, $d_{{obs}}$ is high, THEN Turning is 0.3904;

r6: IF $\theta_{{target}}$ is high, $d_{{target}}$ is low, $d_{{obs}}$ is high, THEN Turning is 0.3938;

r7: IF $\theta_{{target}}$ is high, $d_{{target}}$ is high, $d_{{obs}}$ is high, THEN Turning is 0.5467;

r8: IF $\theta_{{target}}$ is high, $d_{{target}}$ is high, $d_{{obs}}$ is low, THEN Turning is 0.5505;
\\
\hline
\end{tabular}
\label{tab:turtlebot_pmt1}
\end{table*}

\begin{table*}[ht]
\centering
\caption{Prompt of Turtlebot Obstacle Avoidance Case (Part 2)}
\begin{tabular}{p{0.95\textwidth}}
\hline
\textbf{Interpreter Prompt (continued)}:   \\ 
\hline

\#\# Task:

Generate a professional Controller Behavior Explanation Report with the following structure:

$'''$
\\ 
\# Controller Behavior Explanation Report

\#\# Summary of Controller Behavior
Provide a brief overview of the Turtlebot's obstacle avoidance behavior in Case 2 and explain why the navigation process is divided into two explanation phases.

\#\#\# Phase 1: Obstacle Bypassing Analysis

\#\#\#\# State Contribution Analysis

Analyze the explanatory variable significance values table to explain:

- Which explanatory variables have the most significant impact on the Turtlebot's turning behavior in this phase

- The relative importance of each variable

- How the salience values reflect the controller's obstacle avoidance strategy

\#\#\#\# Controller Logic Analysis

Analyze the IF-THEN rules table to explain:

- The key conditions governing the Turtlebot's turning decisions during obstacle bypassing

(Hint: Try to group the rules with similar turning outputs and identify common premises in the IF parts)

- How obstacle-related and target-related variables jointly influence turning behavior

- Why larger or smaller turning actions occur under certain conditions

- The overall control strategy used to bypass obstacles

\#\#\# Phase 2: Target Reorientation Analysis

\#\#\#\# State Contribution Analysis

Analyze the explanatory variable significance values table to explain:

- Which explanatory variables dominate the Turtlebot's turning behavior in this phase

- The relative importance of each variable

- How the salience values reflect the controller's target-reaching strategy

\#\#\#\# Controller Logic Analysis

Analyze the IF-THEN rules table to explain:

- The key conditions governing the Turtlebot's turning decisions after obstacle clearance

(Hint: Try to group the rules with similar turning outputs and identify common premises in the IF parts)

- How target-related and obstacle-related variables jointly influence turning behavior

- Why stronger or weaker turning actions occur under different conditions

- The overall control strategy used to reorient toward the target

\#\#\# Cross-Phase Behavioral Transition Analysis

Compare the explanations of Phase 1 and Phase 2 to explain:

- How the controller's decision-making strategy changes across the two phases

- Whether the controller shifts its priority from obstacle avoidance to target tracking

- How the dominant explanatory variables and rule logic evolve between phases

- What this transition reveals about the controller's overall navigation strategy

\#\#\# Conclusion

Provide a concise summary that:

- Synthesizes the insights from both explanation phases

- Highlights the most critical factors driving the Turtlebot's turning behavior

- Explains the overall obstacle avoidance and target navigation characteristics of the controller

$'''$ \\
\\

\#\# Requirements:

- Use clear, professional English suitable for technical documentation

- Provide specific references to salience values and IF-THEN rules when making explanations

- Ensure the report is comprehensive yet accessible to users who need to understand the controller behavior

- Focus on practical insights that help users understand why and how the Turtlebot controller makes decisions

- Do not merely restate the rules individually, instead summarize the overall control strategy revealed by the rules

- Pay particular attention to explaining behavioral transitions between the two phases

\\ \hline
\end{tabular}
\label{tab:turtlebot_pmt2}
\end{table*}

\begin{table*}[ht]
\centering
\caption{Answer of Turtlebot Obstacle Avoidance Case (Part 1)}
\begin{tabular}{p{0.95\textwidth}}
\hline
\textbf{Interpreter Answer}: \\ 
\hline
\textbf{\normalsize Controller Behavior Explanation Report} \\
\\
\textbf{\normalsize Summary of Controller Behavior}

In this case, the Turtlebot performs obstacle avoidance during the later stage of navigation, when the target lies directly behind an obstacle located along the robot’s forward direction. This configuration creates a navigation conflict: the controller must simultaneously avoid collision with the obstacle while maintaining progress toward the target.

As shown by the segmented trajectory, the avoidance process naturally unfolds in two consecutive behavioral phases. In Phase 1 (Obstacle Bypassing Stage), the Turtlebot actively deviates from the obstacle to establish a safe path around it. During this stage, obstacle positioning and heading alignment dominate the controller’s turning decisions. In Phase 2 (Target Reorientation Stage), after safely bypassing the obstacle, the Turtlebot shifts its attention toward restoring alignment with the target and completing navigation. The explanation results therefore reveal how the controller dynamically changes its decision-making priorities throughout obstacle avoidance.\\
\\
\textbf{\small Phase 1: Obstacle Bypassing Analysis}

\textbf{State Contribution Analysis}

The salience values indicate that angular information dominates the controller’s turning behavior during obstacle bypassing, while distance information plays only a secondary role.

The two most influential variables are:

\begin{itemize}
    \item $\theta_{{obs}}$ (0.0740)
    \item $\theta_{{target}}$ (0.0708)
\end{itemize}

These values are substantially larger than those of:

\begin{itemize}
    \item $d_{{obs}}$ (0.0071)
    \item $d_{{target}}$ (0.0059)
\end{itemize}

This salience distribution suggests that the controller primarily relies on relative directional geometry rather than distance measurements when avoiding the obstacle. Since the target is located directly behind the obstacle, the critical challenge is determining how to steer around the obstacle while maintaining an eventual path toward the target. Consequently, the controller focuses on angular relationships, specifically:

\begin{enumerate}
    \item Obstacle orientation relative to the robot’s heading ($\theta_{\text{obs}}$), which determines collision risk and safe steering direction.
    \item Target orientation relative to the heading ($\theta_{\text{target}}$), which helps preserve navigation toward the final destination.
\end{enumerate}

The near-equal importance of these two angular variables indicates that the controller performs a balanced trade-off between obstacle avoidance and target tracking during bypassing. Rather than reacting only to the obstacle, the controller simultaneously considers where the target lies, enabling smoother and more goal-directed avoidance trajectories.

In contrast, the relatively small contributions of $d_{\text{obs}}$ and $d_{\text{target}}$ imply that precise distances are not the main determinants of turning magnitude in this phase. This is reasonable because the robot has already entered an avoidance maneuver, and the immediate priority becomes directional adjustment, not distance regulation.

\textbf{Controller Logic Analysis}

The IF-THEN rules reveal a structured turning policy primarily governed by the interaction between $\theta_{\text{target}}$ and $\theta_{\text{obs}}$. The rules can be grouped into three turning regimes.

\textbf{1. Weak Turning Behavior}

Rules r1–r3 produce the weakest turning actions (approximately 0.19–0.24):

\begin{itemize}
    \item r1: target angle high, obstacle angle medium → 0.1946
    \item r2: target angle high, obstacle angle high → 0.2166
    \item r3: target angle medium, obstacle angle high → 0.2358
\end{itemize}

These rules share a common pattern: either the obstacle or target is positioned relatively far from the robot’s forward direction (high angular deviation). In such situations, the immediate collision threat is lower or the target is already displaced away from the forward direction, reducing the need for aggressive steering. As a result, the controller applies only moderate corrections.

\textbf{2. Moderate Turning Behavior}

Rules r4–r6 generate intermediate turning outputs (approximately 0.25–0.29):

\begin{itemize}
    \item r4: target angle high, obstacle angle low → 0.2456
    \item r5: target angle medium, obstacle angle medium → 0.2648
    \item r6: target angle low, obstacle angle high → 0.2902
\end{itemize}

These rules indicate situations where either the obstacle or target becomes more behaviorally relevant. For example:

When the obstacle lies close to the forward direction (low $\theta_{\text{obs}}$, r4), stronger turning becomes necessary to initiate bypassing.
When the target lies closer to the forward direction (low $\theta_{\text{target}}$, r6), the controller adjusts turning to preserve eventual target alignment.

This reflects a compromise between maintaining safety and preserving route efficiency.

\textbf{3. Strong Turning Behavior}

Rules r7–r9 produce the strongest turning actions (approximately 0.34–0.36):

\begin{itemize}
    \item r7: target angle low, obstacle angle low → 0.3402
    \item r8: target angle medium, obstacle angle low → 0.3544
    \item r9: target angle low, obstacle angle medium → 0.3604
\end{itemize}

These rules share an important characteristic: the target and obstacle are simultaneously positioned near the forward direction, creating strong directional conflict. This scenario is expected in the current case because the target is located behind the obstacle along nearly the same path.

When both entities compete for the robot’s forward heading, the controller responds with strong turning actions to rapidly establish a safe bypass trajectory. Larger turning outputs therefore represent an active obstacle circumvention strategy rather than simple directional correction.

Overall, the controller adopts a goal-aware avoidance strategy: it avoids the obstacle aggressively when necessary but continuously incorporates target orientation information to avoid excessive detours.

\\ \hline
\end{tabular}
\label{tab:turtlebot_ans1}
\end{table*}

\begin{table*}[ht]
\centering
\caption{Answer of Turtlebot Obstacle Avoidance Case (Part 2)}
\begin{tabular}{p{0.95\textwidth}}
\hline
\textbf{Interpreter Answer (continued)}: \\ 
\hline

\textbf{\small Phase 2: Target Reorientation Analysis}

\textbf{State Contribution Analysis}

After obstacle clearance, the salience pattern changes substantially, indicating a shift in behavioral priorities.

The dominant variables become:

\begin{itemize}
    \item $d_{\text{target}}$ (0.0602)
    \item $\theta_{\text{target}}$ (0.0434)
\end{itemize}

while obstacle-related variables contribute less:

\begin{itemize}
    \item $d_{\text{obs}}$ (0.0198)
    \item $\theta_{\text{obs}}$ (0.0009)
\end{itemize}

The extremely small salience of $\theta_{\text{obs}}$ suggests that the obstacle no longer significantly influences turning behavior, implying successful obstacle clearance.

Instead, the controller transitions toward target-reaching behavior, where the remaining challenge is restoring heading alignment and efficiently approaching the destination. Notably, distance to the target becomes the most important explanatory variable, indicating that turning intensity is influenced not only by directional error but also by how far the robot remains from the target.

The moderate contribution of $d_{\text{obs}}$ suggests that obstacle distance still plays a minor safety-monitoring role, preventing premature abandonment of obstacle awareness immediately after bypassing.

Overall, the salience values reveal a controller that re-prioritizes navigation objectives, shifting from collision avoidance to target acquisition.

\textbf{Controller Logic Analysis}

The rules in Phase 2 reveal a clearer hierarchical decision-making structure centered on target alignment, with turning outputs divided into weak, moderate, and strong regimes.

\textbf{1. Weak Turning Behavior}

Rule r1 produces the smallest turning output (0.2216):

\begin{itemize}
    \item target angle low
    \item target distance low
    \item obstacle distance low
\end{itemize}

This condition corresponds to situations where the robot is already reasonably aligned with and close to the target. Since only minor corrections are needed, turning remains weak.

\textbf{2. Moderate Turning Behavior}

Rules r2–r6 produce medium-strength turning actions (approximately 0.36–0.39):

Examples include:

\begin{itemize}
    \item target angle low, target distance high → 0.3602 (r2)
    \item target angle high, target distance low → 0.3677 (r4)
    \item target angle high, obstacle distance high → 0.3938 (r6)
\end{itemize}

These rules indicate that larger corrections emerge whenever one important navigation factor becomes challenging, such as:

\begin{itemize}
    \item the target remaining far away,
    \item heading misalignment with the target,
    \item or residual obstacle proximity requiring cautious steering.
\end{itemize}

The controller therefore performs meaningful turning corrections whenever efficient reorientation is necessary.

\textbf{3. Strong Turning Behavior}

Rules r7–r8 generate the strongest turning actions (0.5467–0.5505):

\begin{itemize}
    \item target angle high
    \item target distance high
\end{itemize}

regardless of obstacle distance.

This pattern is highly informative: when the robot is both far from the target and poorly aligned with it, the controller consistently issues strong turning commands to rapidly reorient toward the target direction. The fact that obstacle distance no longer significantly changes the output reinforces the conclusion that target tracking has become the dominant control objective.

Overall, the controller adopts a target-driven recovery strategy after obstacle bypassing. Turning intensity increases primarily in response to target misalignment and remaining travel distance, enabling efficient trajectory correction toward the destination.

\\ \hline
\end{tabular}
\label{tab:turtlebot_ans2}
\end{table*}

\begin{table*}[ht]
\centering
\caption{Answer of Turtlebot Obstacle Avoidance Case (Part 3)}
\begin{tabular}{p{0.95\textwidth}}
\hline
\textbf{Interpreter Answer (continued)}: \\ 
\hline

\textbf{\normalsize Cross-Phase Behavioral Transition Analysis}

Comparing the two phases reveals a clear transition in the controller’s decision-making strategy.

\textbf{Shift in Decision Priorities}

In Phase 1, the controller prioritizes safe obstacle bypassing. The dominant explanatory variables are angular quantities, especially $\theta_{\text{obs}}$ (0.0740) and $\theta_{\text{target}}$ (0.0708). This indicates that the controller mainly asks:

- “\textit{How should I change direction to avoid collision while preserving access to the target?}”

In Phase 2, priority shifts toward target reacquisition. The controller instead emphasizes $d_{\text{target}}$ (0.0602) and $\theta_{\text{target}}$ (0.0434), reflecting a new question:

- “\textit{How should I realign efficiently with the target and complete navigation?}”

This transition strongly suggests a dynamic priority switch from obstacle avoidance to target tracking.

\textbf{Evolution of Dominant Variables}

The explanatory variables evolve in a meaningful way:

\begin{itemize}
    \item Obstacle angle importance collapses from 0.0740 → 0.0009, indicating obstacle clearance.
    \item Target distance importance increases from 0.0059 → 0.0602, showing growing emphasis on goal completion.
    \item Target angle remains consistently important, reflecting its persistent role in directional control across both phases.
\end{itemize}

This evolution demonstrates that the controller adaptively reallocates attention depending on the navigation context.

\textbf{Evolution of Rule Logic}

The IF-THEN rules also change structurally:

\begin{itemize}
    \item Phase 1 rules emphasize competition between obstacle and target orientation, producing turning behaviors designed to establish a collision-free path.
    \item Phase 2 rules become target-centric, where turning intensity mainly depends on target distance and angular misalignment.
\end{itemize}

In Phase 1, strong turning emerges when target and obstacle directions conflict. In Phase 2, strong turning occurs when the robot remains far from and poorly aligned with the target.

This rule transition reflects a controller that behaves strategically rather than reactively.

\textbf{Implications for Overall Navigation Strategy}

Taken together, the explanations suggest that the controller implements a two-stage navigation policy:

\begin{enumerate}
    \item Safety-first obstacle circumvention, where turning behavior is governed by directional conflict between obstacle and target.
    \item Goal-directed trajectory recovery, where turning behavior focuses on restoring efficient movement toward the target.
\end{enumerate}

This demonstrates adaptive and context-sensitive decision-making throughout navigation.

\textbf{\normalsize Conclusion}

The explanation results reveal that the Turtlebot controller exhibits a clear and interpretable two-phase navigation strategy during late-stage obstacle avoidance.

During Phase 1, turning behavior is primarily driven by angular relationships with both the obstacle and target, enabling the robot to safely bypass the obstacle while maintaining awareness of the destination. Strong turning actions occur when the target and obstacle occupy similar forward directions, creating navigational conflict.

During Phase 2, the controller shifts toward target-oriented reorientation, where distance and alignment to the target become the dominant factors governing turning behavior. Obstacle-related influence decreases substantially, confirming successful obstacle clearance.

Overall, the controller demonstrates an adaptive balance between collision avoidance and target pursuit, dynamically reallocating attention as navigation conditions change. The most critical factors driving turning behavior are therefore directional geometry during obstacle bypassing and target distance/alignment after clearance, together enabling safe and efficient navigation.

\\ \hline
\end{tabular}
\label{tab:turtlebot_ans3}
\end{table*}

\bibliographystyle{ieeetr}  
\bibliography{ref}